\documentclass[acmsmall,letterpaper]{acmart}

\AtBeginDocument{%
  \providecommand\BibTeX{{%
    \normalfont B\kern-0.5em{\scshape i\kern-0.25em b}\kern-0.8em\TeX}}}

\setcopyright{acmcopyright}
\copyrightyear{2020}
\acmYear{2020}

\renewcommand\footnotetextcopyrightpermission[1]{} %

\usepackage{framed}

\usepackage{multirow}
\usepackage{booktabs}
\usepackage{ifthen}
\usepackage{color}
\usepackage{url}
\usepackage{amsmath}
\usepackage{xspace}
\usepackage[T1]{fontenc}
\usepackage{enumerate}
\usepackage[linesnumbered,ruled,vlined]{algorithm2e}
\usepackage{array,multirow,graphicx}
\usepackage{float}
\usepackage{balance}
\usepackage{tikz}
\usepackage{calc}
\usepackage{subfigure}
\usepackage{listings,amsfonts}
\usepackage{amsmath,xcolor,pifont}
\usepackage{url}
\usepackage{xspace}
\usepackage{endnotes}
\usepackage{xcolor}
\usepackage{array}
\usepackage{multirow,makecell}
\usepackage[misc]{ifsym}
\usepackage{graphicx}
\usepackage{hyperref}

\definecolor{yellow}{RGB}{255,255,153}
\definecolor{grey}{RGB}{224,224,224}

\begin{document}
\title{Ethereum Name Service: the Good, the Bad, and the Ugly}

\author{Pengcheng Xia}
\affiliation{%
  \institution{Beijing University of Posts and Telecommunications}
  \city{Beijing}
  \country{China}
}

\author{Haoyu Wang}
\authornote{Corresponding Author: Haoyu Wang (haoyuwang@bupt.edu.cn).}
\affiliation{%
  \institution{Beijing University of Posts and Telecommunications}
  \city{Beijing}
  \country{China}
}
\email{haoyuwang@bupt.edu.cn}

\author{Zhou Yu}
\affiliation{%
  \institution{Beijing University of Posts and Telecommunications}
  \city{Beijing}
  \country{China}
}

\author{Xinyu Liu}
\affiliation{%
  \institution{Beijing University of Posts and Telecommunications}
  \city{Beijing}
  \country{China}
}

\author{Xiapu Luo}
\affiliation{%
  \institution{The Hong Kong Polytechnic University}
  \city{HongKong}
  \country{China}
}

\author{Guoai Xu}
\affiliation{%
  \institution{Beijing University of Posts and Telecommunications}
  \city{Beijing}
  \country{China}
}

\renewcommand{\shortauthors}{Pengcheng Xia et al.}

\begin{abstract}
DNS has always been criticized for its inherent design flaws, making the system vulnerable to kinds of attacks. Besides, DNS domain names are not fully controlled by the users, which can be easily taken down by the authorities and registrars. Since blockchain has its unique properties like immutability and decentralization, it seems to be promising to build a decentralized name service on blockchain. Ethereum Name Service (ENS), as a novel name service built atop Etheruem, has received great attention from the community. Yet, no existing work has systematically studied this emerging system, especially the security issues and misbehaviors in ENS. 
To fill the void, we present the first large-scale study of ENS by collecting and analyzing millions of event logs related to ENS. We characterize the ENS system from a number of perspectives. Our findings suggest that ENS is showing gradually popularity during its four years' evolution, mainly due to its distributed and open nature that ENS domain names can be set to any kinds of records, even censored and malicious contents. We have identified several security issues and misbehaviors including traditional DNS security issues and new issues introduced by ENS smart contracts. Attackers are abusing the system with thousands of squatting ENS names, a number of scam blockchain addresses and malicious websites, etc. Our exploration suggests that our community should invest more effort into the detection and mitigation of issues in Blockchain-based Name Services towards building an open and trustworthy name service.

\end{abstract}

\maketitle

\section{Introduction}

The domain name system (DNS) has already become an indispensable component of the functionality of the Internet since 1980s. It operates like a phone book, i.e., translating human-readable domain names to the numerical IP addresses for accessing other computers or devices under these addresses. DNS is built in a hierarchical and distributed way, which makes the system resilient and scalable. However, DNS has always been criticized for its inherent design flaws, making the system vulnerable to kinds of attacks~\cite{dnsvul}. 
A recent report suggested that a group of hackers have launched DNS hijacking attack at least 30 organizations, including government ministries, embassies and security services as well as companies and other groups in Europe and the Middle East~\cite{reutersreport}.

Thus, many efforts were made from both the research community and the industry, aiming to make the DNS system more secure and reliable~\cite{doh,dot,dnssec}.
For example, the Domain Name System Security Extensions (DNSSEC) is one of the attempts~\cite{dnssec}. DNSSEC is a set of extensions to DNS to support cryptographic authentication of DNS data, authenticated denial of existence, and data integrity. 
However, DNSSEC does not provide confidentiality of data, which means that DNSSEC responses are authenticated but not encrypted. Besides, the signing and checking process of the digital signature will increase the DNS query latency and affect the user experience. Moreover, the complexity of implementing and maintaining DNSSEC impedes its adoption, making it hard to be widely deployed in the wild~\cite{dnssecdis}. 

With the prosperity of blockchain techniques in recent years, researchers are exploring ways to address the issues of traditional DNS by combing blockchain with DNS. 
Since blockchain has its unique properties like immutability and decentralization, it seems to be promising to build a decentralized name service atop blockchain. Indeed, some blockchain-based DNS (BNS) solutions have already been proposed. For example, Namecoin~\cite{namecoin} is claimed to be the first blockchain-based DNS solution, which is a fork of Bitcoin network that offers a new \texttt{.bit} top-level domain (TLD) for its names. 
Similar to Namecoin, UnstoppableDomains~\cite{unstoppable} and EmerDNS~\cite{emerdns} proposed new TLDs like \texttt{.zil} , \texttt{.crypto} and \texttt{.emc}, etc. 
Handshake~\cite{handshake} takes another way, in which it attempts to replace the DNS root with a more decentralized, secure system. The names on these BNS services can be exclusively owned and managed by their owners, which cannot be taken by the authority (e.g., the government) and can be more affordable to some extent. However, every coin has two sides. 
The advantages of BNS can be exploited to fulfill malicious purposes. According to a recent study~\cite{huang2020leopard}, blockchain domain names have been exploited as command and control (C\&C) channels by attackers. Furthermore, the domain squatting issues~\cite{squatting} on the BNS services will be more severe than the normal DNS, due to the difficulty of shutting down malicious/phishing blockchain domains.

Ethereum Name Service (ENS)~\cite{enshome} is a decentralized naming service built atop Ethereum~\cite{ethereum}, one of the most popular blockchains that allows users to create dApps (Decentralized Applications) by developing smart contracts.
Different from the aforementioned BNS solutions, ENS aims to propose a complementary solution to DNS by taking advantage of smart contracts on Ethereum to manage the registration and resolution of domain names. 
It focuses first and foremost on resolving names to \texttt{web3} resources like blockchain addresses and decentralized websites (dWebs). 
According to the official announcement~\cite{enshome}, it has been integrated by more than 170 popular services including blockchain wallets, dApps and browsers. For example, browsers like Chrome and Firefox have extensions to resolve ENS domain names when users type them into browsers directly. Therefore, ENS has become one of the most popular blockchain name systems in the wild. %

Although ENS has been deployed for roughly 4 years, to our best knowledge, it has not been systematically studied in our research community. We are still unaware of the status quo of this emerging name service, especially the security issues and the dark side of the system. There remain a number of unexplored questions, e.g., \textit{how many domain names are registered in ENS? how people are using ENS? whether security issues and gray behaviors are prevalent in ENS?}

\textbf{This work.} In this paper, we take the first step to systematically characterize ENS. 
To fully understand the registration and resolution process, we first fetch and analyze all the event logs of ENS-related smart contracts and third party resolver contracts (See \textbf{Section~\ref{sec:studydesign}}). At last, we get over 5.6 million event logs. By decoding these logs, we harvest $465,827$ registered names and $107,617$ Ethereum addresses that ever used ENS. 
Based on this dataset, we perform a detailed analysis of each operating period on ENS (See \textbf{Section~\ref{sec:general}}). We observe that 39.3\% of all ENS names are active and 74.4\% of users are active (i.e., at least have one name) by the time of our study. We then explore the usage of ENS names by decoding the ENS records and observe that over 67\% of record settings are related to blockchain addresses (See \textbf{Section~\ref{sec:records}}).
To further examine the security issues of ENS, we have adopted a series of measurement studies (See \textbf{Section~\ref{sec:security}}) to investigate both the traditional security issues (i.e., domain name squatting, malicious domains and scam addresses) and the ENS specific security issue (i.e., name record persistence attack).
We obtain the following key findings:

\begin{itemize}
    \item \textbf{ENS is showing gradually popularity during its four years' evolution.} Over 465K ENS names were registered, and 180K of them are active by the time of this study. A number of users are willing to pay high prices for rare ENS names or get as many names as they can. 

    \item \textbf{ENS is a fully open system where the ENS domain names can be set to any kinds of records, and ENS is on its way to becoming a complementary system of DNS.} The most common use of ENS name is to link to blockchain addresses, accounting for 67.3\% of the record settings. Besides, it is also popular to use ENS names for decentralized websites. We also find that people are exploring new ways to interacting with ENS through text records.
    
    \item \textbf{The open nature of ENS makes it easy to be abused by attackers.} We have identified several security issues and misbehaviors, including traditional DNS security issues and new issues introduced by ENS smart contracts. A few squatters are found holding a lot of famous brand names and their variants, which could be used for malicious purposes. Some malicious decentralized websites and scam addresses are found in records of ENS names. Besides, record persistence issue is found that may cause potential attacks.
\end{itemize}

To the best of our knowledge, this is the first comprehensive study of Ethereum Name Service, one of the most promising BNS solutions at scale, longitudinally and across various dimensions. Our results motivate the need for more research efforts to illuminate the widely unexplored BNS systems. We believe that our efforts can attract the focus of the research community and promote best operational practices across BNS solutions. We will release our dataset, along with the experiment results to the research community (link removed due to anonymous submission).

\section{Background}

\begin{figure}[t]
\centering
\includegraphics[width=0.95\textwidth]{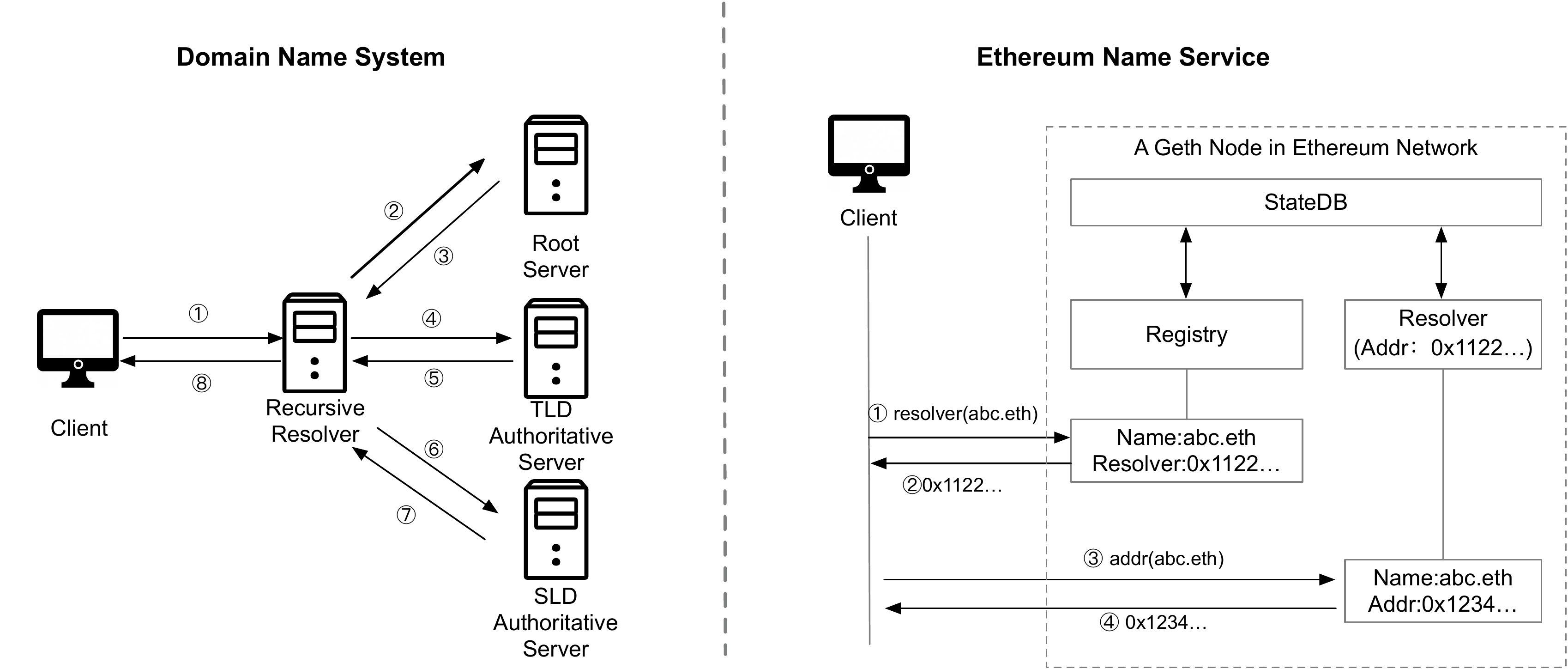}
\vspace{-0.1in}
\caption{A comparison of the structures of DNS and ENS.}
\vspace{-0.2in}
\label{fig:dnsensstruct}
\end{figure}

\subsection{Domain Name System and Zooko's Triangle}

\subsubsection{Domain Name System} 
Domain Name System is a network protocol that associates domain names with various network information. The most common use of DNS is to map domain names to IP addresses, which will help different kinds of computers and other network resources connect to the Internet or private networks. The DNS mapping is distributed throughout the Internet in a hierarchy authority.
A typical DNS resolution process is shown in Figure~\ref{fig:dnsensstruct}. When a client wants to query for the IP of a domain name, it will first query the recursive resolver. 
If the resolver has the proper information, it will return the information directly. Otherwise, it will query the root servers which has the information of top-level domains (TLDs), and get the information of the relevant TLD server. Then, a query will be made to TLD server for second-level domain (SLD) authoritative server. After receiving the request, the SLD server could respond the corresponding IP address to the resolver. The whole DNS lookup process can take milliseconds.

However, DNS has been criticized for many years, due to various kinds of vulnerabilities and security issues. 
For example, due to the lack of authentication and integrity checking in DNS, attacks like DNS cache poisoning and DNS tunneling were prevalent in the network~\cite{dnsvul}. Besides, DNS domain names are not fully controlled by the users and can be easily taken down by the authorities and registrars. Some efforts like DNS over HTTPS (DoH), DNS over TLS (DoT) and DNSSEC are made to solve or alleviate some of these issues~\cite{doh,dot,dnssec}. 
People also use other solutions like Tor or Invisible Internet Project (I2P) for privacy~\cite{tor,i2p}. Nevertheless, DNS and the aforementioned solutions still cannot achieve \textit{human-readability}, \textit{security} and \textit{decentralization} simultaneously, which is known as Zooko's Triangle.

\subsubsection{Zooko's Triangle}
Zooko's Triangle~\cite{zooko} defines three properties that an ideal name system should possess: 
1) \textit{human-meaningful}, i.e., the names should be readable and can be memorable by human;
2) \textit{secure}, i.e., the names should be translated correctly even when the system is attacked; 
3) \textit{decentralised}, i.e., the names should be translated without central authority in the system. The proposer of this triangle, Zooko Wilcox-O'Hearn, also speculated that any name system could only achieve two of these three properties at most. 
This triangle has been used to evaluate a name system's performance. For example, DNS names are human-readable but the DNS system is not secure and decentralized. Tor is secure and decentralized but its addresses are not human-readable. I2P addresses are human-readable and secure, but I2P is not decentralized. 
With the rising of blockchain techniques, some blockchain-based name systems have been proposed, which are claimed to fulfill all three properties in Zooko's Triangle~\cite{namecoin,emerdns,unstoppable,handshake}.

\subsection{Blockchain Name Service and Ethereum Name Service}

\subsubsection{Blockchain Name Service (BNS)}
Blockchain techniques have raised great concerns since Bitcoin was invented in 2009, leading to the revolution in many fields. Since blockchain has its unique properties like immutability and decentralization, it seems to be promising to build a decentralized name service on blockchain. Therefore, some blockchain-based name services are proposed in recent years. The most common purpose of BNS is to replace the traditional DNS with blockchain-based alternatives. They usually propose new TLDs that are incompatible with the traditional DNS and the Internet Corporation for Assigned Names and Numbers (ICANN). For example, Namecoin~\cite{namecoin} is the first blockchain-based DNS, which claims to be the first solution of Zooko's Triangle. 
It proposes the \texttt{.bit} TLD for users and has the ability to attach identity information or human-meaningful Tor domains. Similarly, UnstoppableDomains~\cite{unstoppable} and EmerDNS~\cite{emerdns} also propose some new TLDs. Besides, Handshake~\cite{handshake} is another kind of BNS, which seeks to replace the DNS root with its decentralized, secure system.

\begin{table}[t]
\caption{The 8 type of records in the public resolvers.}
\vspace{-0.1in}
\resizebox{0.8\linewidth}{!}{

\begin{tabular}{@{}ll@{}}
\toprule
Record Type & Description                \\ \midrule
Address       & Can be ETH address or other blockchain address                                \\
Name          & Used for reverse resolution, i.e., mapping wallet addresses to ENS names      \\
Content Hash  & IPFS hash,  Swarm hash for dWebs and Tor .onion address hash          \\
Text          & Key-value text record, and key can be "email",   "URL", "vnd.twitter", etc. \\
DNS Record  & DNS record in wire-format  \\
Pubkey      & ECDSA SECP256k1 public key \\
ABI           & Application Binary Interface for interacting with contracts                   \\
Authorisation & Granting one address full access to one name except authorisations            \\ \bottomrule
\end{tabular}
}
\vspace{-0.1in}
\label{tab:recordtype}
\end{table}

\subsubsection{Ethereum Name Service (ENS)}
\label{sec:ensback}
Unlike the above BNS attempts, ENS aims to be a complementary solution to DNS by integrating with some traditional TLDs. ENS is built atop Ethereum, the first blockchain system that supports smart contracts. ENS is controlled by several smart contracts, which can interact with users to register and manage names automatically. Specifically, ENS mainly consists of three kinds of smart contracts: the registry, the registrars and the resolvers~\cite{ensdoc}. 

(1) \textit{Registry} stores the mapping of ENS names (of any level) to owners, resolvers and caching time-to-live (TTL) for ENS names' records. In order to avoid trivial enumerations of names during the initial auction (see Section~\ref{sec:vickrey}) and force names with different lengths to be an identifier with fixed length in smart contracts, ENS stores names in the form of hashes of them, which are generated through the process named ``namehash''. The namehash can be calculated by combining the hash of the highest-level part of ENS domain names (called ``labelhash'') with namehash of other part and then performing a hash again on it \footnote{namehash(example.eth) = keccak256(keccak256(example) + namehash(eth))}. This algorithm will preserve the hierarchical properties of ENS names. We will describe how we restore ENS names from hash values in Section~\ref{sec:dataset}. 

(2) \textit{Registrar} is a kind of smart contract that owns a name, which can automatically assign subdomain names to users based on some rules (e.g., payment). 
ENS team has ever used some registrar contracts for \texttt{.eth} name registrations, including the Vickrey auction registrar and the permanent registrar. Besides, along with the permanent registrar, the concept of registrar controller was introduced to delegate the name management of name owners. We will detail the registration processes in the Section~\ref{sec:stage}. 
In particular, users whose DNS names are supported by ENS can claim their DNS names in ENS by proving the ownership through DNSSEC and setting the TXT records containing their Ethereum addresses~\cite{ensxyz}. TLDs like \texttt{.kred} and \texttt{.luxe} can be linked with owners' Ethereum accounts directly in their DNS registrars~\cite{ensluxe}.

(3) \textit{Resolver} stores the mapping of names to records. ENS can store arbitrary records while the ``public resolvers'' implemented by ENS team have predefined eight types of records (see Table~\ref{tab:recordtype}).

As shown in Figure~\ref{fig:dnsensstruct}, the ENS name resolution is a two-step process. The user who wants to resolve the name needs to query the registry to find the correct resolver and then get the resolution results from the resolver.

\section{The Evolution of Ethereum Name Service}
\label{sec:stage}

\begin{figure}[t]
\centering
\includegraphics[width=0.95\textwidth]{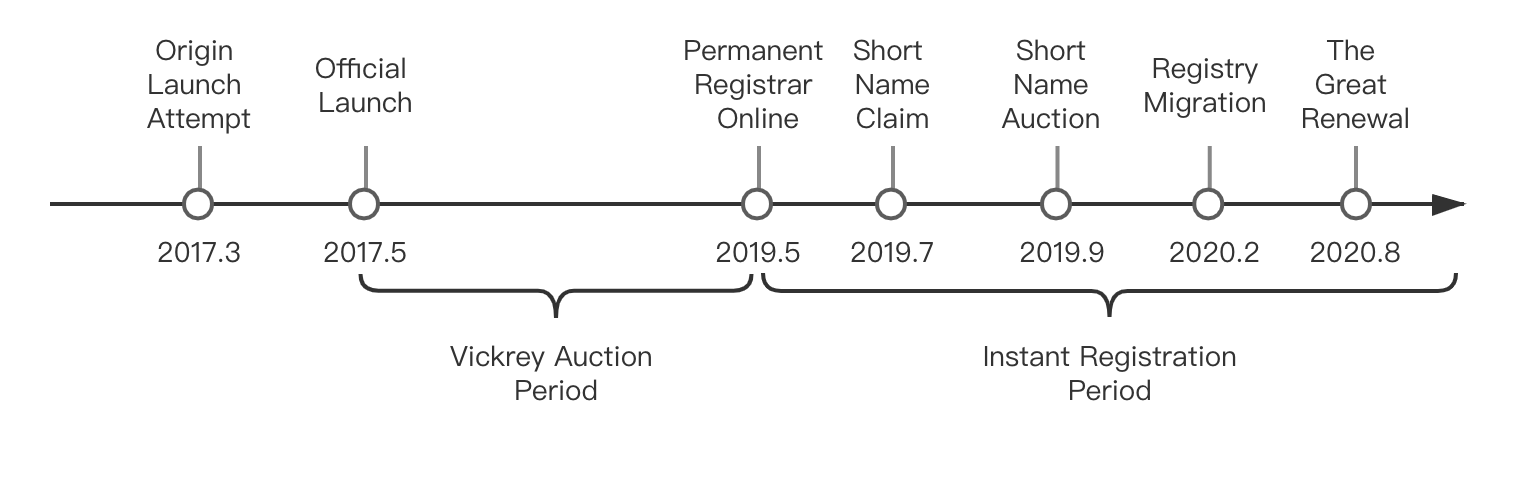}
\vspace{-0.4in}
\caption{The timeline of major ENS events.}
\vspace{-0.2in}
\label{fig:enstimeline}
\end{figure}

As the registration mechanism of ENS is evolving from time to time, in this section, we will detail how ENS works and how ENS names are registered during each stage. According to the official ENS blogs~\cite{ensblog}, we have summarized the timeline of ENS evolution, as shown in Figure~\ref{fig:enstimeline}. 
The ENS team initially launched the ENS service in 2017 March while they encountered two severe bugs and the service went offline soon after the launch~\cite{ensbug}. 
After the code revision, they re-launched the service on May 4th 2017, which adopted a Vickrey auction to register names.

\subsection{The Initial Auction (Vickrey Auction)}
\label{sec:vickrey}
When ENS formally launched on May 4th 2017, the ENS team deployed a smart contract\footnote{Address:0x6090a6e47849629b7245dfa1ca21d94cd15878ef (Etherscan label: ``ENS: Old Registrar'')} implementing a Vickrey auction for registering names that have a length of more than 6. \textit{Vickrey auction}~\cite{vickrey} is a type of sealed-bid auction where bidders submit their bids without knowing how much others have bid and the winner of the auction is the highest bidder while he only needs to pay the second-highest price. Besides, in ENS Vickrey auction, \texttt{.eth} names are transferred into hashes (as depicted in Section~\ref{sec:ensback}) to avoid trivial enumeration and names would be gradually released during an 8-week period. These schemes to some extent help people have a greater probability of obtaining the names they want. The Ether paid by a name's bidders will be deposited into a smart contract called ``deed'' and all the losers of the auction will get a less 0.5\% refund\footnote{The deed contract would burn 0.5\% of the paid Ether in order to create a cost for large amount registration and for registering valuable names}. The winner of the name only needs to pay the second-highest price, and she could give up the ownership to withdraw all the Ether she paid after registration for one year. We will perform a detailed analysis on the behavior of this Vickrey auction period (see Section~\ref{sec:phase1}).

\subsection{Permanent Registrar, Short Name Claim and Short Name Auction}

\subsubsection{Permanent Registrar}
After two years of auction, ENS team launched the ``permanent registrar'' for registering names over 6 in length instead of the auction registrar on May 4th 2019. The permanent registrar aims to run continuously until the registrar contract has to be replaced due to severe deficiency. 
The charging method of \texttt{.eth} names is changed to an annual rent payment model, in which each name needs to be charged 5 US Dollars per year. 
Along with the permanent registrar, the concept of registrar controller was introduced to delegate the name management of name owners. Thus, a name registered by the registrar controller can set resolver and name records within the registration transaction, which simplifies the registration process.

\subsubsection{The Short Name Claim} 
In July 2019, ENS team opened the reservation of short \texttt{.eth} names (names with a length of 3-6), which means that owners of eligible traditional TLD names can request for corresponding \texttt{.eth} names and pay rent in advance to obtain the access to their corresponding \texttt{.eth} names for one year (\$640 in ETH for a 3 character name, \$160 for a 4 character name, \$5 for a 5-6 character name). An owner of a short second-level traditional name registered on or before May 4th 2019 can claim one of the following names: 1) An exact match of the original name (e.g., \texttt{foo.com} to \texttt{foo.eth}). 2) Removing the \texttt{eth} suffix of original name (e.g., \texttt{fooeth.com} to \texttt{foo.eth}). 3) Combining the 2LD and TLD of the original name (\texttt{foo.com} to \texttt{foocom.eth}). Upon application, the ENS team will review the request for validity. 

\subsubsection{The Short Name Auction} 
In September 2019, another auction for remaining short names with a length of 3-6 started. 
The ENS team chose OpenSea~\cite{opensea}, a well-known crypto assets marketplace, as the auction platform, and used \textit{English auction}~\cite{english} as the auction method. In an English auction, bids are public and bidders can bid multiple times. The bidder who submits the highest price will win the name and the payment he deposited will be the registration fee of the first year, which is quite different from the Vickrey auction period. After the short name auction, the remaining short names will be open for registration at a price based on their length. According to analysis in Section~\ref{sec:phase2}, there were few DNS name owners claiming corresponding ENS names while many famous brand names are selling high price in short name auction. This could be related to squatting behaviors and we will further investigate it in Section~\ref{sec:popreg}.

\subsection{The Great Renewal}
Since ENS introduced the ``permanent registrar'', expiration and renewal mechanisms like traditional domain names were also introduced into this decentralized naming service. Currently in ENS, all the \texttt{.eth} names are charged annually based on their name length and anyone can renew any \texttt{.eth} names no matter whether they own the name or not. For old names registered through the Vickrey auction, they were set to expire on May 4th 2020 if not renewed. Besides, all the \texttt{.eth} names will have a 90-day grace period after expiration where payment can be made to keep the ownership. 

Thus, due to a large number of names were registered in the Vickrey auction period, most of the names will be expired on May 4th 2020 (actually expired on August 2nd considering the 90-days grace period) and the renewal period around August 2020 was called ``The Great Renewal''~\cite{ensgreat}. To avoid massive squatting behaviors and gas competition for registration priority, the ENS team carried out the ``decaying price premium''~\cite{ensdecay}, where the price of expired names will start at \$2,000 combining with normal annual rent and will decrease linearly to normal annual rent in 28 days.

\section{Study Design}
\label{sec:studydesign}

We present the details of our measurement study on ENS in this section. We first describe the research questions, and then present how we collect the ENS data used for our study.

\subsection{Research Questions}

Our study aims to understand the status quo of the ENS ecosystem, and investigate the security issues. To this end, our study is driven by the following research questions (RQs).

\begin{itemize}
    \item [RQ1] \textbf{Popularity of ENS.} Considering ENS has been launched over 4 years and its purpose is to be complementary to DNS, it is thus necessary to investigate its popularity, i.e., \textit{how many domain names are registered, and how many addresses are involved in ENS?} The results could shed light on the adoption level of ENS in the community, which will be studied in Section~\ref{sec:general}.

    \item [RQ2] \textbf{Usage of ENS.}
    Considering the unique features provided by ENS, it is unknown to us \textit{how people are using ENS except for the explicit address translation}. Therefore, it is interesting to analyze the records of all the ENS names. This RQ will be answered in Section~\ref{sec:records}.
    
    \item [RQ3] \textbf{Security Issues of ENS.}
    Since no existing work has systematically analyzed the security issues in the ENS ecosystem, it is important to understand \textit{whether security issues (both traditional DNS security issues and new emerging ENS specific issues) are prevalent in ENS, and how severe are them}. The observations could offer implications for the design of blockchain-based name services. We will explore the security issues and misbehaviors in Section~\ref{sec:security}.

\end{itemize}

\begin{figure}[h]
\centering
\includegraphics[width=0.9\textwidth]{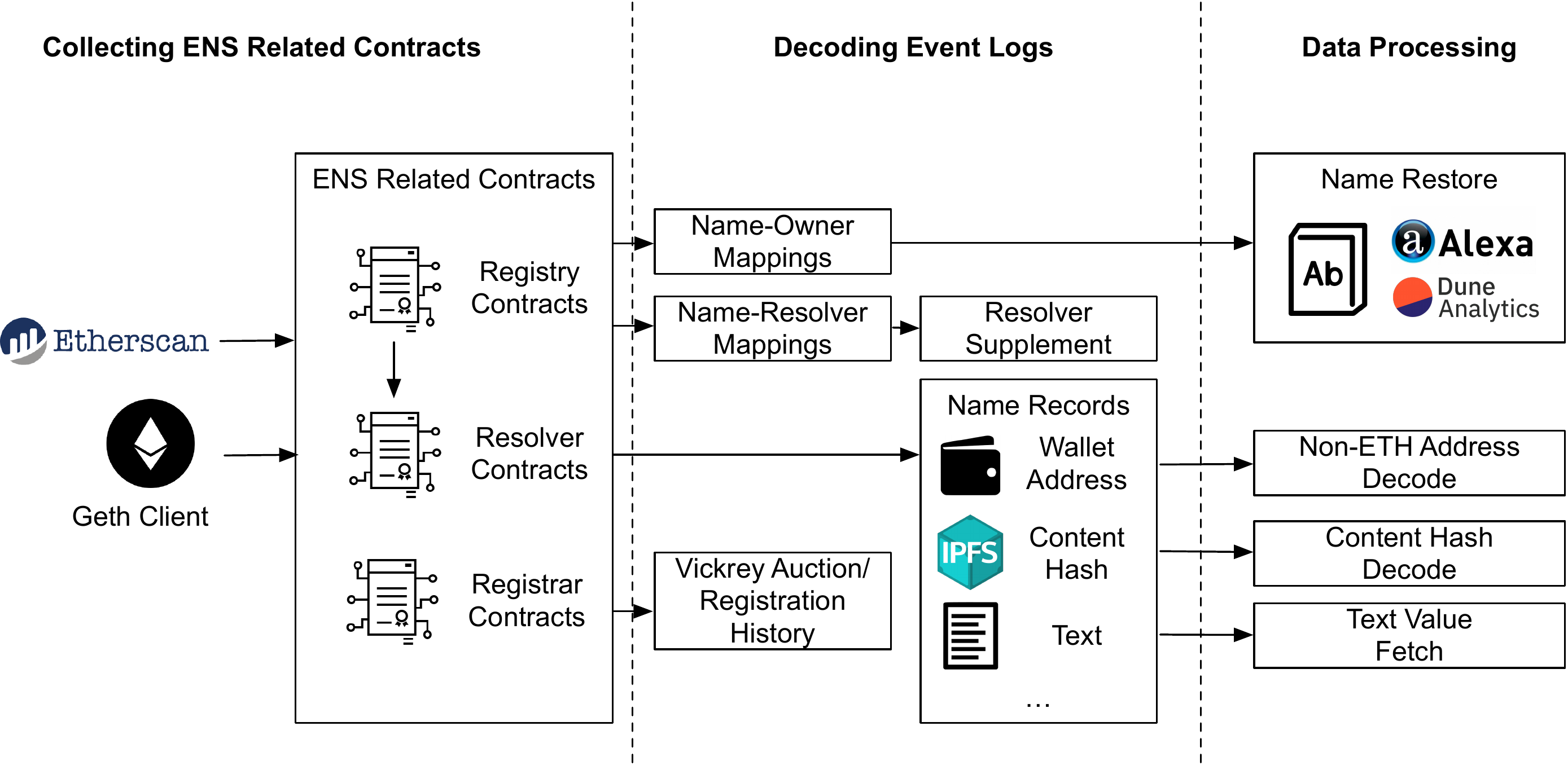}
\vspace{-0.1in}
\caption{The workflow of our data collection.}
\vspace{-0.2in}
\label{fig:workflow}
\end{figure}

\subsection{Dataset Collection}
\label{sec:dataset}

It is non-trivial to harvest all kinds of information related to ENS. First, ENS stores ENS domain names (ENS names for short, the same below) in the form of hash values so that we cannot get their human-readable names directly. Second, ENS has multiple resolvers (including third-party resolvers) and they use different types of protocols to encode records in ENS names.
Thus, we follow a hybrid workflow to extract comprehensive information on ENS, which is shown in Figure~\ref{fig:workflow}. Our dataset collection contains three major steps.

\subsubsection{Collecting ENS-related smart contracts.} 
The first step is to collect all ENS official smart contracts that are highly related to core functions of ENS (e.g., name registration and name renewal). Thus, we resort to Etherscan~\cite{etherscan}, a commonly used Ethereum explorer, to search for related contracts. Etherscan has labeled 28 ENS official smart contracts with human-meaningful names. For example, the contract used for Vickrey auction\footnote{Address: 0x6090A6e47849629b7245Dfa1Ca21D94cd15878Ef} is labeled as ``Old Registrar''. 
Note that some smart contracts are not related to the core functionalities of ENS, e.g., the ``Multisig'' contract is used for effect administrative changes. Thus, we only focus on the aforementioned three types of smart contracts that are related to the resolution of ENS, including registry contracts, resolver contracts, registrar contracts (including registrar controller contracts and a short name claim contract). We manually analyzed all these contracts and labelled 14 of them.

\subsubsection{Decoding Event Logs.} 
After collecting related contracts, we take advantage of Geth~\cite{geth}, a well-known Ethereum client to synchronize the ledger of Ethereum. Specifically, in order to get the state changes of each contract, we extract event logs\footnote{Event logs will record the major activities of smart contracts and this will help track smart contracts' behaviors.} from the ledger. Then, since ENS official contracts are open-sourced on Etherscan, we fetch the Application Binary Interface (ABI) \footnote{ABI encodes how to interact with the functions of a smart contract.} of each contract and decode event logs based on their ABIs. Thus, we get name-owner mappings, name-resolver mappings from registry contracts, obtain name records history from resolver contracts, and collect auction/registration history from registrar contracts. Furthermore, in name-resolver mappings, we find that lots of names are pointed to additional resolvers. Thus, we further make effort to include open-source extra resolvers, fetch their event logs, and decode them based on their ABIs, which is shown in Table~\ref{tab:exresolver} of Appendix.

\begin{table}[t]
\caption{An overview of the ENS event logs we collected.}
\vspace{-0.1in}
\small
\resizebox{0.9\linewidth}{!}{
\begin{tabular}{@{}|ccc|r|@{}}
\toprule
\begin{tabular}[c]{@{}c@{}}Contract \\ Type\end{tabular} &
  Etherscan Name Tag &
  Address &
  \multicolumn{1}{c|}{\begin{tabular}[c]{@{}c@{}}\# of \\ event logs\end{tabular}} \\ \midrule
\multicolumn{1}{|c|}{\multirow{2}{*}{Registry}} & Eth Name Service               & 0x314159265dD8dbb310642f98f50C066173C1259b & 1,097,965 \\
\multicolumn{1}{|c|}{}                          & Registry with Fallback         & 0x00000000000C2E074eC69A0dFb2997BA6C7d2e1e & 944,486   \\ \midrule
\multicolumn{1}{|c|}{\multirow{3}{*}{Registrar}} &
  Base Registrar Implementation &
  0x57f1887a8BF19b14fC0dF6Fd9B2acc9Af147eA85 &
  1,135,019 \\
\multicolumn{1}{|c|}{}                          & Old ENS Token                  & 0xFaC7BEA255a6990f749363002136aF6556b31e04 & 337,071   \\
\multicolumn{1}{|c|}{}                          & Old Registrar                  & 0x6090A6e47849629b7245Dfa1Ca21D94cd15878Ef & 1,979,316 \\
\multicolumn{1}{|c|}{}                          & Short Name Claims                  & 0xf7C83Bd0c50e7A72b55a39FE0DABF5e3A330d749 & 883 \\\midrule
\multicolumn{1}{|c|}{\multirow{3}{*}{\begin{tabular}[c]{@{}c@{}}Registrar \\ Controller\end{tabular}}} &
  Old ETH Registrar Controller 1 &
  0xF0AD5cAd05e10572EfcEB849f6Ff0c68f9700455 &
  14,976 \\
\multicolumn{1}{|c|}{}                          & Old ETH Registrar Controller 2 & 0xB22c1C159d12461EA124b0deb4b5b93020E6Ad16 & 20,827    \\
\multicolumn{1}{|c|}{}                          & ETHRegistrarController         & 0x283Af0B28c62C092C9727F1Ee09c02CA627EB7F5 & 57,483    \\ \midrule
\multicolumn{1}{|c|}{\multirow{5}{*}{Resolver}} & OldPublicResolver1             & 0x1da022710dF5002339274AaDEe8D58218e9D6AB5 & 14,997    \\
\multicolumn{1}{|c|}{}                          & OldPublicResolver2             & 0x226159d592E2b063810a10Ebf6dcbADA94Ed68b8 & 19,510    \\
\multicolumn{1}{|c|}{}                          & PublicResolver1                & 0xDaaF96c344f63131acadD0Ea35170E7892d3dfBA & 2,603     \\
\multicolumn{1}{|c|}{}                          & PublicResolver2                & 0x4976fb03C32e5B8cfe2b6cCB31c09Ba78EBaBa41 & 61,410    \\
\multicolumn{1}{|c|}{}                          & Additional Resolvers           & --                                         & 110,869   \\ \bottomrule
\end{tabular}
}
\vspace{-0.1in}
\label{tab:eventlog}
\end{table}

\subsubsection{Data Processing.} 

In the decoded event logs, some information like Vickrey auction history can be directly used for further research. 
However, some additional data processing tasks are still needed due to the design of these contracts. 
Specifically, we need to get the unhashed ENS names from name-owner mappings, decode non-ETH wallet addresses and content hashes based on their encoding rules and fetch the values of text records from corresponding transactions.

As stated in Section~\ref{sec:ensback}, ENS smart contracts store hash values of ENS names instead of names themselves, thus we take efforts to restore these hash values to readable names. 
First, ENS developers uploaded their name-hash dictionary to Dune Analytics~\cite{duneanalytics}, a platform for querying, extracting and visualizing Ethereum data. We fetch the dictionary from the website and update our database based on it. Moreover, we generate the labelhash based on a list of over 460K English words and Alexa~\cite{alexatop} top-100K domain list (downloaded on Sept. 18th 2020) and match them with the hashes in registry event logs (this data is also used in Section~\ref{sec:popreg} for identifying squatting names). Besides, the ``NameRegistered'' and ``NameRenew'' events of new ENS registrar controllers contain the plain texts of newly registered names and we simply add them to our database.

For the name records, since non-ETH addresses and content hashes have been encoded for uniformity, we decode them based on the rules in EIP-2304~\cite{eip2304} and EIP-1577~\cite{eip1577}\footnote{Hashes in old resolvers' ``ContentChanged'' events are treated as Swarm hashes as they did not have a uniformed format.}. For the text records, as stated in EIP-634~\cite{eip634} and in ENS docs~\cite{ensdoc}, the event logs of them only contain the keys but not the values. Thus, we resort to the transaction data related to these event logs from the Ethereum ledger and decode them based on ABIs to get the text values.

\subsection{Dataset Overview}
The dataset collection is a complicated and time-consuming process, since we need to cover all the cases to create a comprehensive dataset. 
At last, we get all the ledger information until block $10,746,639$ (i.e., 2020-08-28 03:03:42 UTC) on Ethereum. 
The overall statistics is shown in Table~\ref{tab:eventlog}. 
In total, we get over 2 million registry logs, 3.4 million registrar logs, and 200 thousand resolver logs.
In addition to event logs, we also fetch and decode over $3,000$ transactions related to text records. 
We find $465,827$ ENS names in the registry records\footnote{We exclude ENS top-level domain names' (TLDs) records and reverse resolution names because our study is focused on second and higher level ENS names.}.
Besides, we restore 373,950 names (including 323,255 \texttt{.eth} names, which accounts for 86.6\% of all \texttt{.eth} names) in total. To the best of our knowledge, it is the largest preimage ENS name dataset by the time of the study, which is even larger than the dataset provided by the official team.

\section{General Overview of Ethereum Name Service}
\label{sec:general}
In this section, we will first depict the general overview of ENS, and then perform a deep analysis on each phase during the evolution of ENS.

\subsection{Overall Statistics}
\label{subsec:overall}

\begin{table}[h]
\small
\caption{The distribution of ENS names. Note that, the .eth subdomain owners of expired parent names and expired DNS name owners still have control over their ENS names, which are considered as active ENS names.}
\vspace{-0.1in}
\begin{tabular}{@{}|cr|cr|@{}}
\toprule
Category                      & \multicolumn{1}{c|}{\# of Names} & Category                   & \multicolumn{1}{c|}{\# of Names} \\ \midrule
\texttt{.xyz }   Names  & 96    & Total ENS Names          & 465,827 \\
\texttt{.club} Names & 1     & \texttt{.eth} Names           & 373,241 \\
\texttt{.luxe} Names & 1,914 & Unexpired \texttt{.eth} Names & 90,576  \\
\texttt{.art }   Names  & 27    & Expired \texttt{.eth} Names   & 282,665 \\
\texttt{.kred} Names & 216   & \texttt{.eth} Subdomain Names         & 90,332  \\ \midrule
\textbf{DNS Integrated Names} & \textbf{2,254}                   & \textbf{ENS Names by Study Time} & \textbf{183,162}                 \\ \bottomrule
\end{tabular}
\vspace{-0.2in}
\label{tab:namesoverall}
\end{table}

\subsubsection{Overview}
Table~\ref{tab:namesoverall} shows an overview of the $465,827$ ENS names. $107,617$ addresses have ever participated in the registration of ENS \texttt{.eth} names, and by the study time there are over 183K active names, which are related to 74.4\% of addresses ($80,081$)\footnote{The owner data is different from ENS home page, which uses the data created by the official in DuneAnalytics. We have checked the source code on it and find its problem, which is confirmed by ENS official team. }. Besides \texttt{.eth} names, there are also $2,254$ DNS names\footnote{ENS introduced traditional DNS TLD \texttt{.ceo} on August 26th 2020, and it has no related names in our dataset.} involved in the registration of ENS, which to some extent shows one of the ENS's vision of being complementary of DNS service~\cite{ensintegrate}.

\subsubsection{The evolution of ENS names.}
\begin{figure}[t]
\centering
\includegraphics[width=0.85\textwidth]{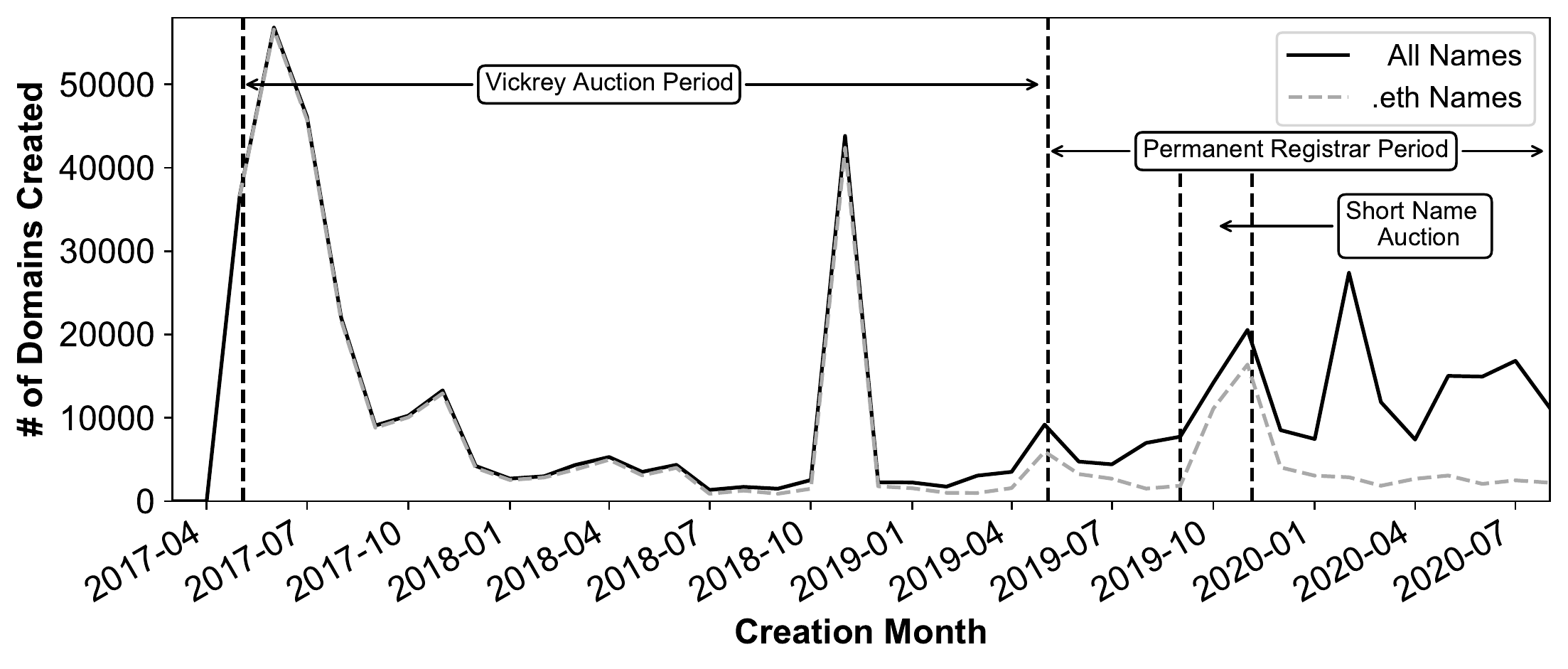}
\vspace{-0.15in}
\caption{The monthly distribution of ENS names' registrations.}
\vspace{-0.15in}
\label{fig:ensreg}
\end{figure}

Figure~\ref{fig:ensreg} shows the monthly registration. We use the first block time of ``NewOwner'' event of names as their first registration time, and the figure shows the trend of ENS names (all ENS names and \texttt{.eth} names) registered for the first time each month.
The ENS team once started their service in March 2017, but they encountered two severe bugs and the service went offline~\cite{ensbug}. Thus, only basic ENS names like \texttt{.eth} or \texttt{addr.reverse} were registered on March 2017 and were deregistered soon. On May 4th 2017, ENS relaunched and the first ENS name registered on May 9th (after a 5-day auction period) is \texttt{rilxxlir.eth}. The first 7 months after the launch witnessed people's enthusiasm for holding ENS \texttt{.eth} names, when $192,471$ names (51.6\% of all \texttt{.eth} names) were registered. 
There is a peak in November 2018, when $43,832$ were registered. In this month, 4 addresses registered a large number of Chinese pinyin names (e.g., \texttt{tianxian.eth}) and names composed of date or numbers (e.g., \texttt{20140409.eth}), which resulted in them ranking 2-5 in the registration numbers of the auction period. 
On May 4th 2019, the ENS team launched a new registration registrar instead of the old auction process. The number of registrations increased slightly until the short name auction started from September to November. The short name auction also affected the registration of other names in October and November\footnote{The auction period lasted more than 1 month.}. In February, Decentraland, a decentralized virtual reality platform, created over 12K subdomain names for their own naming system~\cite{dcl}, which led to a rise in the numbers of names.

\subsubsection{The owners of ENS names.} 
We track every events related to ownership changes of \texttt{.eth} names from ENS registry (i.e., ``NewOwner'' and ``Transfer'' events) and analyze the number of \texttt{.eth} names held by each addresses. 
These 373K ENS \texttt{.eth} names were ever owned by $42,154$ Ethereum addresses. 
Over 35\% of the addresses have more than one names by the time of this study, indicating that although ENS now has an annual fee mechanism on \texttt{.eth} names, there are still a large amount of users holding numerous \texttt{.eth} names. The top-10 name owners by the time of this study are shown in Table~\ref{tab:top10owner} in Appendix. These top-10 holders have roughly 10\% of current registered \texttt{.eth} names and it can be inferred that these top holders tend to keep the names they registered for future benefits. The address that holds most names is  \texttt{0xbcbd4885ee8b2b74249c5ad9b8b668fb256a51b1} with 2,262 names, which registered many words in dictionary (e.g., \texttt{plead.eth} and \texttt{height.eth}) and famous brands (e.g.,  \texttt{disneyplus.eth}). This address and some other top holders are suspicious to involve in squatting behaviors, which will be analyzed in Section~\ref{sec:security}.

\subsubsection{The length of names.}
\begin{figure}[t]
\begin{tabular}{cc}
\begin{minipage}[t]{0.44\linewidth}
    \includegraphics[width = 1\linewidth]{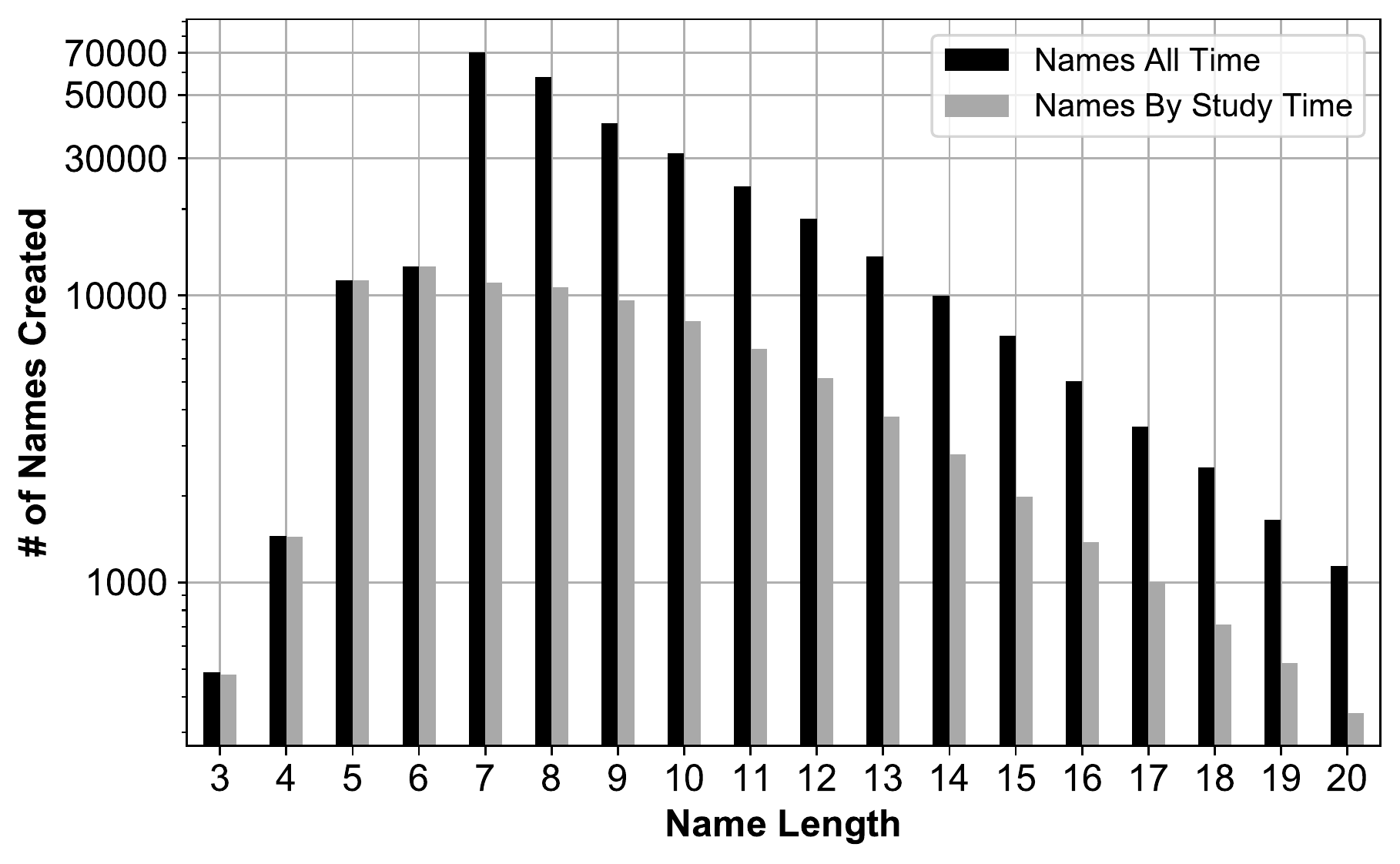}
    \caption{The distribution of \texttt{.eth} names' lengths.}
    \label{fig:namelen}
\end{minipage}
\begin{minipage}[t]{0.48\linewidth}
    \includegraphics[width = 1\linewidth]{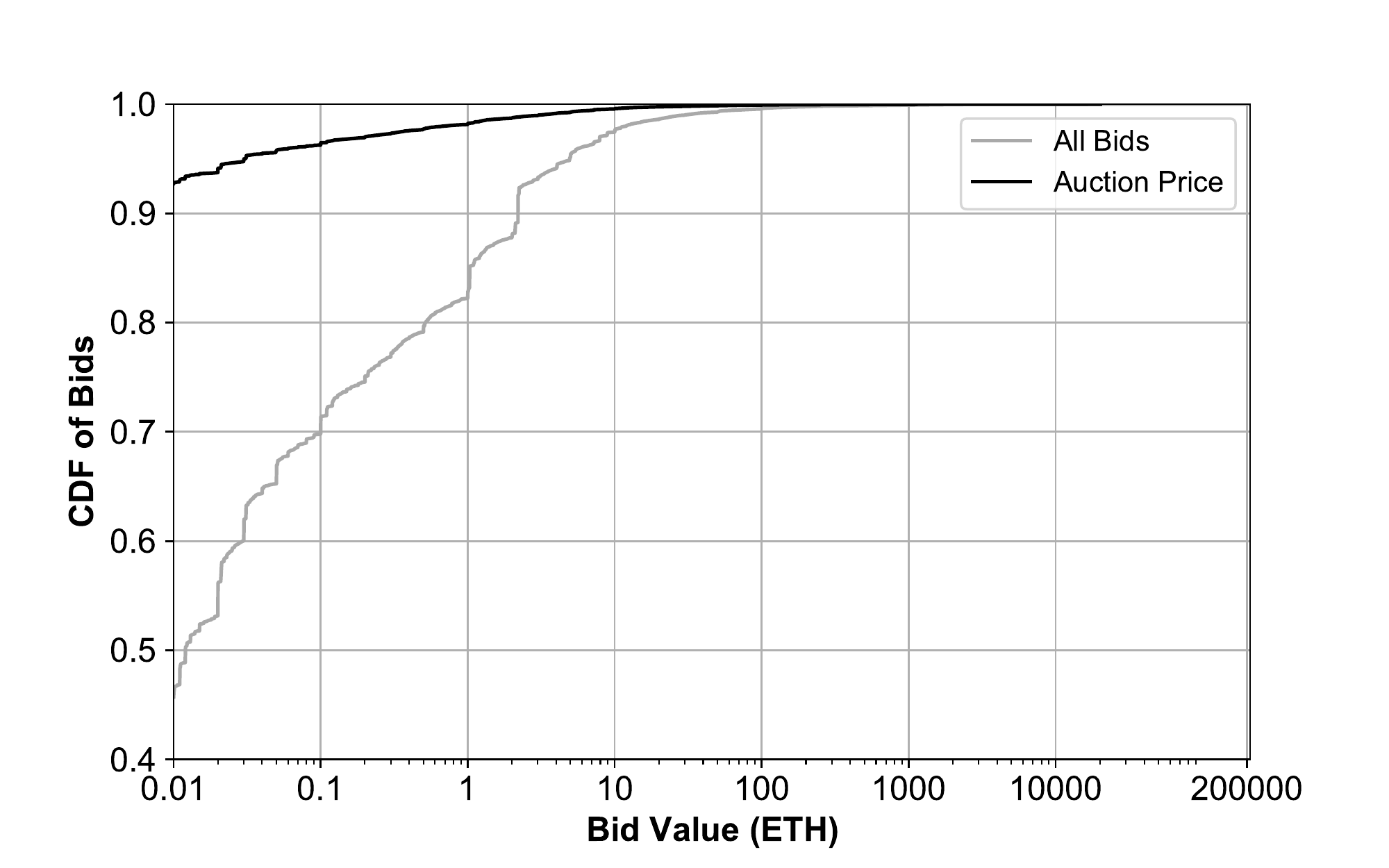}
    \caption{The distribution of bids and auction price.}
    \label{fig:initauctioncdf}
\end{minipage}
\end{tabular}
\vspace{-0.2in}
\end{figure}

We further analyze the popularity of \texttt{.eth} names of different lengths with restored ENS names. Figure~\ref{fig:namelen} shows the distribution of \texttt{.eth} names whose length is under 20. ENS initially only accepted name registration with a length larger than 6 and opened shorter name registration after claim and auction. Besides, currently names with a length of less than 5 will be charged more than \$160 annually. This explains why \texttt{.eth} names that have a length of larger than 6 are more popular. By the time of the study, names with a length ranging from 5 to 8 account for 50.4\% of unexpired \texttt{.eth} names, which are the major choices of registrations. There have been $2,705$ \texttt{.eth} names whose length is over 20 and the longest name has 242 characters of ``a''.

\subsection{The Initial Auction (Vickrey Auction)}
\label{sec:phase1}

\subsubsection{Overview of Vickrey auction.} 
In this period, there were $361,751$ names that have been bid. Among them, $274,052$ names were registered with $338,252$ valid bids by $17,625$ addresses. Note that over 80K names did not finish the auction process or the auctions were cancelled due to their short name length found by other users. 
Figure~\ref{fig:initauctioncdf} shows the distribution of the bids and the final auction price. Interestingly, 45.7\% of the bids are 0.01 ETH while 92.8\% of the auction names valued 0.01 ETH. The highest bid was $201,709$ ETH for \texttt{ethfinex.eth} while its auction price is 0.01 ETH. 

\subsubsection{The most valuable names.} 
The top-10 valuable names in the Vickrey auction are shown in Table~\ref{tab:top10auction} of Appendix. The most valuable name is \texttt{darkmarket.eth}, which cost the winner over 20K ETH (about \$ 5 million). This owner of the name\footnote{0x8759b0b1d9cba80e3836228dfb982abaa2c48b97} also registered another three valuable names including \texttt{openmarket.eth}, \texttt{tickets.eth}, and \texttt{payment.eth}. The owner is said to be an address of a well-known cryptocurrency exchange, Bitfinex~\cite{bitfinex}.
Note that 7 of these names had not set any records by the time of this study, indicating that they may be used for squatting purposes.

\subsubsection{The holders involved in ENS Vickrey auction.} 
Table~\ref{tab:top10spent} in Appendix shows the top-10 name holder addresses and top-10 addresses that spent the most during this period, which do not overlap with each other at all. For name holders, each of the top-10 name holders registered over $2,000$ names, while last 9 of them each spent under 150 ETH in total and the top name holder \textit{0xa7f3659c5382\-0346176f7e0e350780df304db179} ranked 15 among the top spent addresses. For top-10 spent addresses, they each spent over $2,500$ ETH on relatively few names. 
This suggests that there are two straightforward strategies when people bid for desired ENS names. Some people tended to register as many names as they can with low prices while some other people preferred to bidding for few high-value names that worth lots of money.

\subsection{Permanent Registrar, Short Name Claim And Short Name Auction}
\label{sec:phase2}

\subsubsection{The Short Name Claim} 
During this period, 344 requests were submitted and 193 were approved. Among the applications, some famous traditional sites like \texttt{nba.com}, \texttt{paypal.cn}, \texttt{ebay.net}, \texttt{opera.com} and \texttt{infura.io} applied and got the corresponding \texttt{.eth} names, indicating that ENS has received attentions beyond the blockchain community. 

\subsubsection{The Short Name Auction} 
Since this auction took place in OpenSea and the details of this auction are not shown in the ENS contracts' event logs, we take advantage of the data shared by OpenSea in ENS blog~\cite{ensshort} to analyze the trends of the auction this time. 
In total, there are over 50K bids and $7,670$ names were sold for $5,697$ ETH during this auction. The price distribution is shown in Figure~\ref{fig:snameprice}. Roughly 10\% of the names have a price of over 1.5 ETH (about \$594 by study time) and over 22\% of the names were bid for over 10 times. The top-10 popular names and expensive names are shown in Table~\ref{tab:top10shortname}. It is not surprising to see that famous companies like ``apple'', ``google'', ``amazon'' and terms related to sex like ``sex'' and ``porn'' are in the popular name list, and we can also find that some blockchain-related ENS names like ``assets'' and ``dapp'' have also become the hot pursuit of people. Compared with the name price in the Vickrey auction period, the name price in the short name auction tend to be relatively low since users need to pay the bids actually instead of depositing the payments in the deeds. Considering that there were few brands claiming their corresponding \texttt{.eth} names in the short name claim period, it is possible that bad actors bid for famous brand names and use them for malicious purposes. We will further investigate whether there are some squatters targeting at ENS in Section~\ref{sec:squatting}.

\begin{table}[]
\small
\caption{The top-10 popular names and expensive names during the short name auction.}
\vspace{-0.1in}
\begin{tabular}{@{}|ccc|ccc|@{}}
\toprule
Name &
  \begin{tabular}[c]{@{}c@{}}\# of \\ Bids\end{tabular} &
  \begin{tabular}[c]{@{}c@{}}Price in \\ ETH\end{tabular} &
  Name &
  \begin{tabular}[c]{@{}c@{}}\# of \\ Bids\end{tabular} &
  \begin{tabular}[c]{@{}c@{}}Price in \\ ETH\end{tabular} \\ \midrule
amazon & 36 & 100  & asset  & 83 & 30   \\
wallet & 51 & 75   & banker & 78 & 10.5 \\
google & 47 & 52.9 & durex  & 70 & 1.4  \\
apple  & 67 & 51   & apple  & 67 & 51   \\
sex    & 44 & 41   & lawyer & 66 & 7.1  \\
porn   & 44 & 40   & hotel  & 60 & 20   \\
com    & 16 & 39.8 & pussy  & 58 & 8    \\
dapp   & 34 & 38.7 & kering & 58 & 1.4  \\
loan   & 30 & 38   & foster & 58 & 1.1  \\
jobs   & 22 & 35.4 & poker  & 57 & 33.5 \\ \bottomrule
\end{tabular}
\vspace{-0.2in}
\label{tab:top10shortname}
\end{table}

\subsection{The Great Renewal}
\label{sec:phase3}

\begin{figure}[t]
\begin{tabular}{ccc}
\begin{minipage}[t]{0.32\linewidth}
     \includegraphics[width=1\linewidth]{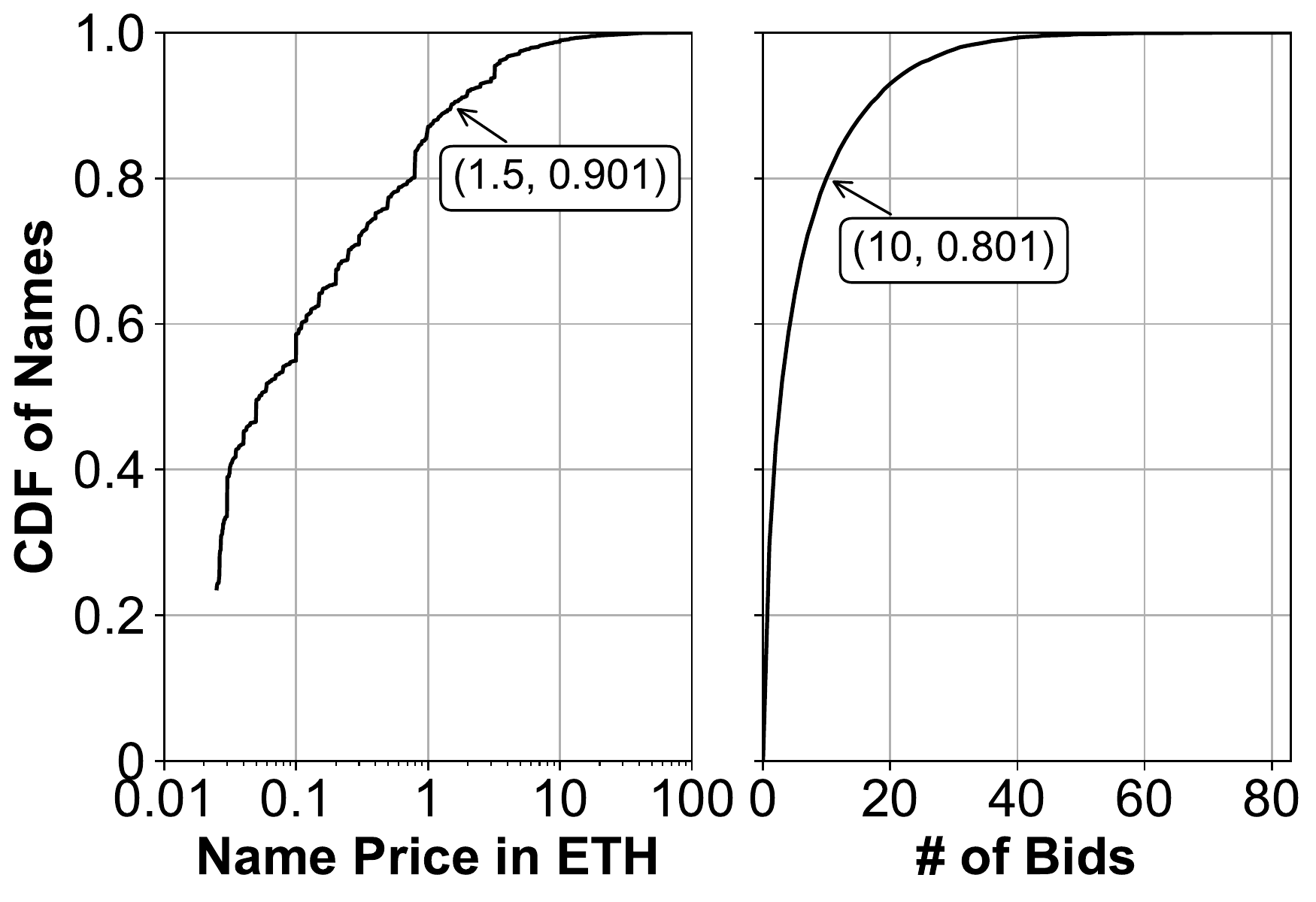}
    \caption{The distribution of short names' price and bids. 
    }
    \label{fig:snameprice}
\end{minipage}
\hspace{0.05in}
\begin{minipage}[t]{0.32\linewidth}
     \includegraphics[width=0.95\linewidth]{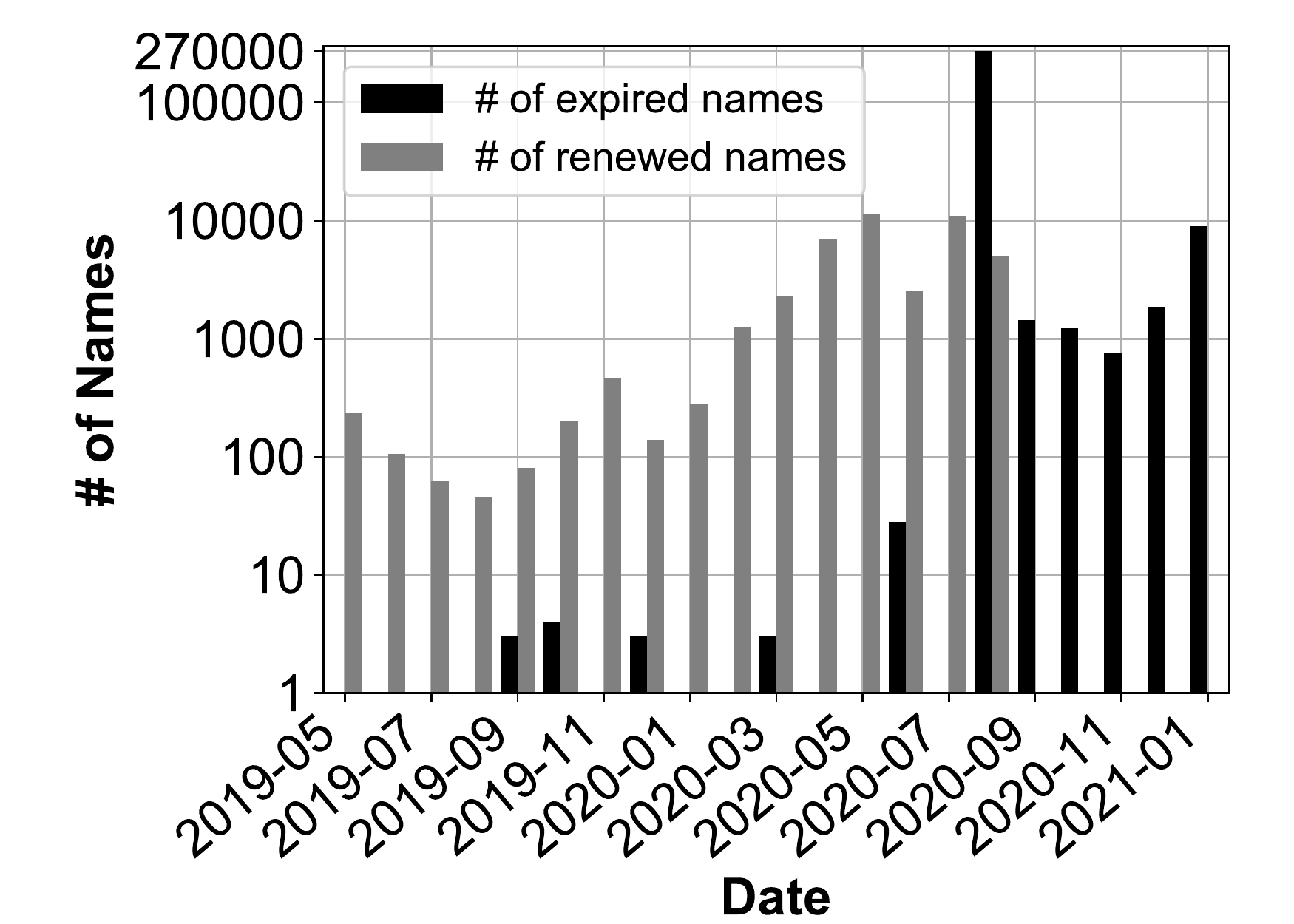}
     \caption{The distribution of expired names and renewed names.}
     \label{fig:exrename}
\end{minipage}
\hspace{0.05in}
\begin{minipage}[t]{0.32\linewidth}
    \includegraphics[width=1\textwidth]{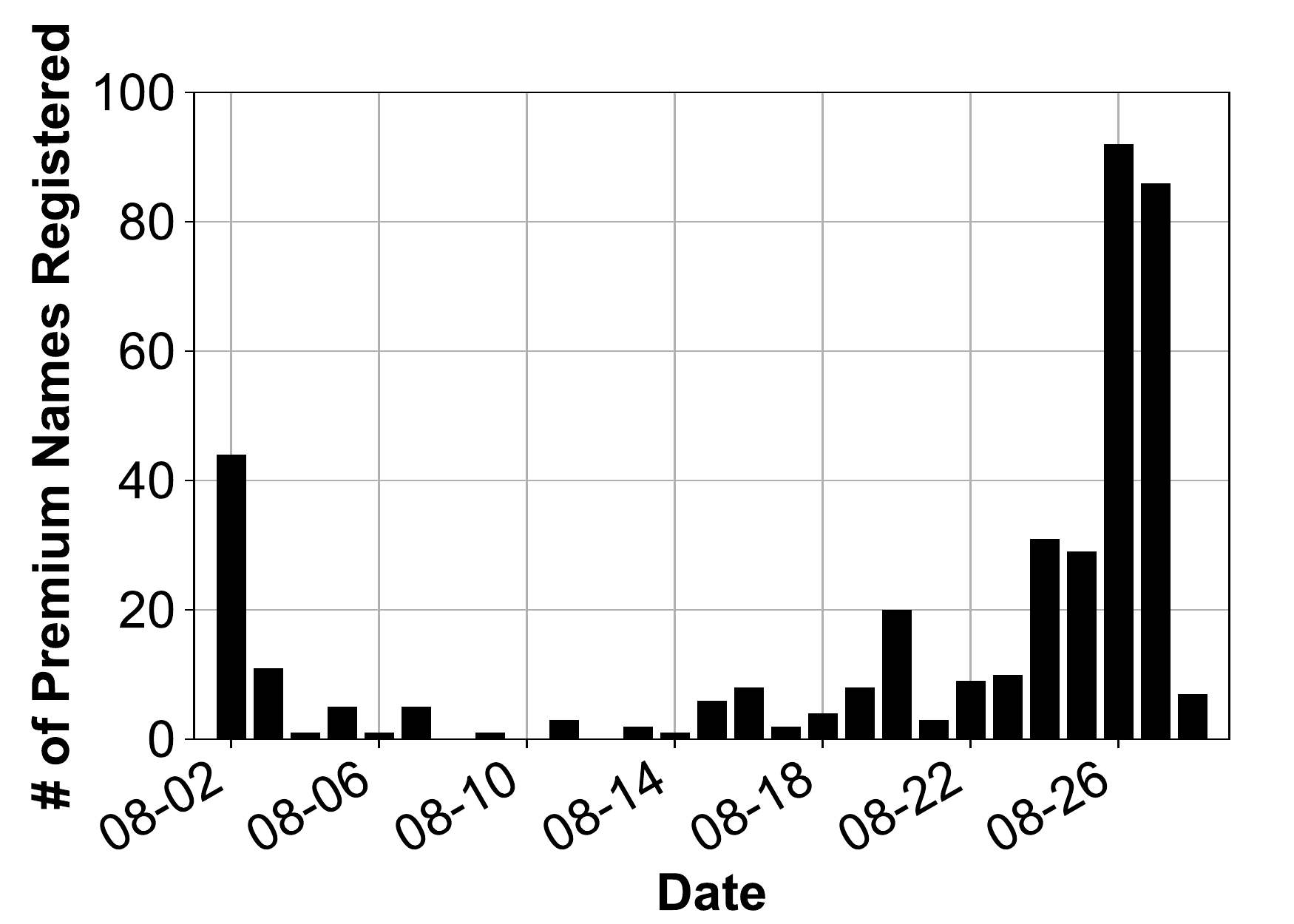}
    \caption{The distribution of premium name registration.}
    \label{fig:premium}
\end{minipage}
\end{tabular}
\vspace{-0.2in}
\end{figure}

Figure~\ref{fig:exrename} shows the distribution of expired names and renewed names (status by the time of the study). Note that we take the 90-day grace period into consideration. It can be seen that, most of the names were expired in August 2020 and the renewals occurred mainly around August 2020. 

Figure~\ref{fig:premium} shows the distribution of 389 premium name registration. There were 44 premium names registered on the first day (August 2nd), suggesting that they were registered with almost full premium. For example, decentralised finance (DeFi) related ENS names like \texttt{makerdao.eth} and \texttt{balancer.eth} were registered almost once upon these ENS names released. 
As design, the first batch of names will be available for registration without premium on August 30th, and that is the reason why we see a spike around the end of August. To some extent, this premium method gives people a greater chance of obtaining the names they strive.

\begin{framed}
\noindent \textbf{Answer to RQ1:} 
\textit{
ENS is showing gradually popularity during its four years' evolution. Over 465K ENS names were registered and 180K of them are active by the time of this study. A number of users are willing to pay high prices for rare ENS names or get as many names as they can. 
}
\end{framed}

\section{The Records of ENS Names}
\label{sec:records}
Next, we analyze how people are using ENS names. As aforementioned, besides linking to Blockchain addresses, ENS also provides functions for users to upload text records, web3 content hash, etc.

\subsection{Overview of ENS records}
After decoding the fetched resolver event logs, we find that over 140K names have been set records over 170K times. Figure~\ref{fig:ensrecord} shows the distribution of record settings. It can be seen that the most widespread use of ENS is as alternatives for blockchain addresses, which accounts for over 67.3\% of total record changes. Other records include content hash records, public key records text records, etc. The distribution of names that have records is shown in Table~\ref{tab:namerecords}. It is interesting to see that only 22\% of the names have ever had records and \texttt{.eth} names, only 12\% (31\% unexpired) names have records.
In ENS, an address can set multiple records. Specifically, the records could at most contain different blockchain addresses, text records with different keywords and other single records. Table~\ref{tab:namerecords} also shows the distribution of the record counts per name. Most names have one record and 98\% of records are Ethereum addresses. The names that have most records are \texttt{brantly.eth} and \texttt{brantly.xyz}, which are names of one of ENS developers and are each set 9 text records, one content hash record, and 15 blockchain addresses records.

\begin{table}[]
\caption{The distribution of names that have records and records per name.}
\vspace{-0.1in}
\begin{tabular}{@{}|cr|cr|@{}}
\toprule
Name Type &
  \multicolumn{1}{c|}{\begin{tabular}[c]{@{}c@{}}\# of \\ Names\end{tabular}} &
  \begin{tabular}[c]{@{}c@{}}\# of Records \\ Per Name\end{tabular} &
  \multicolumn{1}{c|}{\begin{tabular}[c]{@{}c@{}}\# of \\ Names\end{tabular}} \\ \midrule
Name that has records                & 102,557 & 1      & 93,361 \\
\texttt{.eth} name that has records           & 44,297  & 2      & 8,353  \\
Unexpired \texttt{.eth} name that has records & 27,812  & 3 - 25 & 843    \\ \bottomrule
\end{tabular}
\vspace{-0.2in}
\label{tab:namerecords}
\end{table}

\subsection{The use of Blockchain address records.} 
\begin{figure}[htbp]
\centering
\subfigure[The distribution of ENS names' records]{
\includegraphics[width=0.22\linewidth]{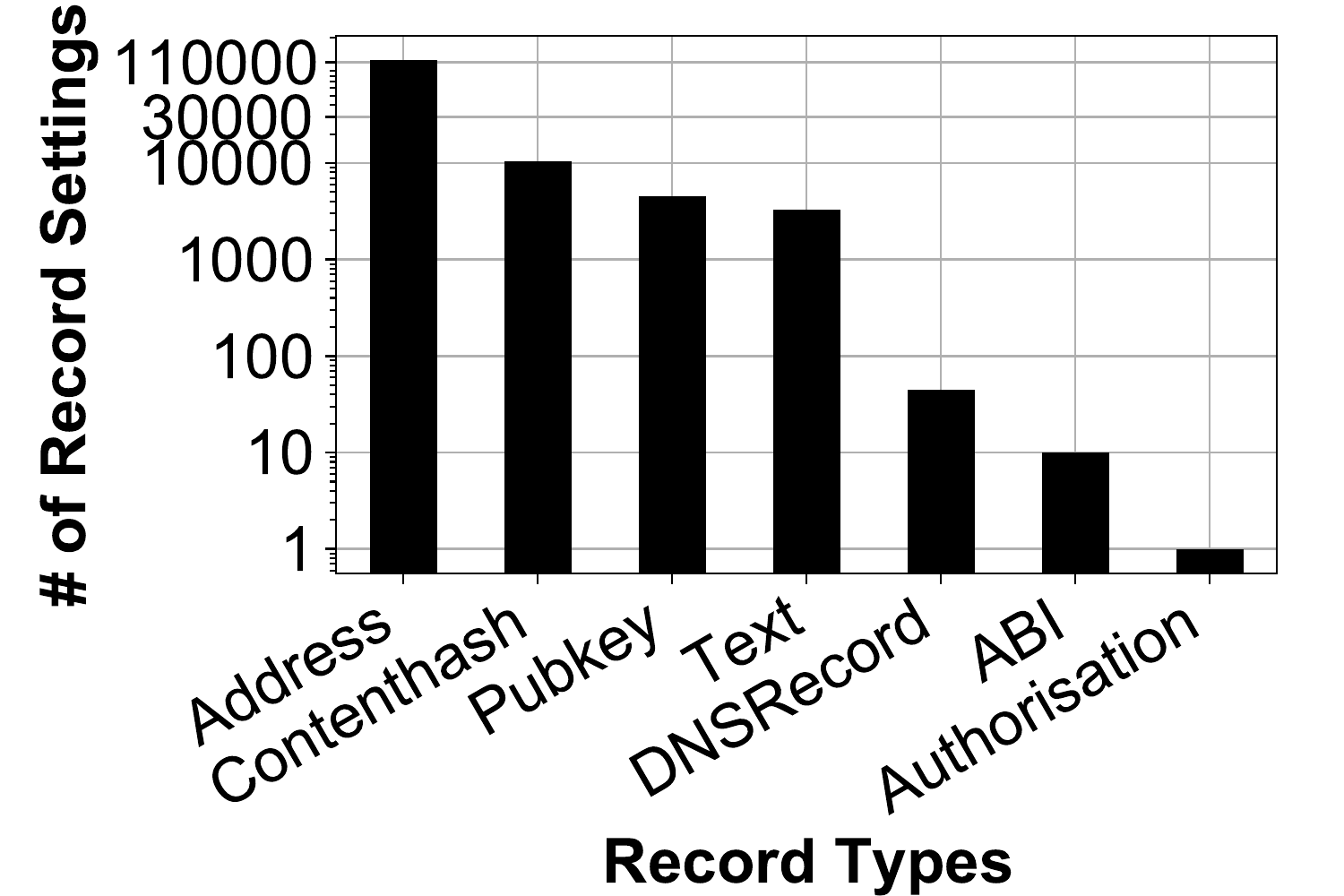}
\label{fig:ensrecord}
}
\hspace{0.05in}
\subfigure[The distribution of top-5 non-ETH addresses.]{
\includegraphics[width=0.22\linewidth]{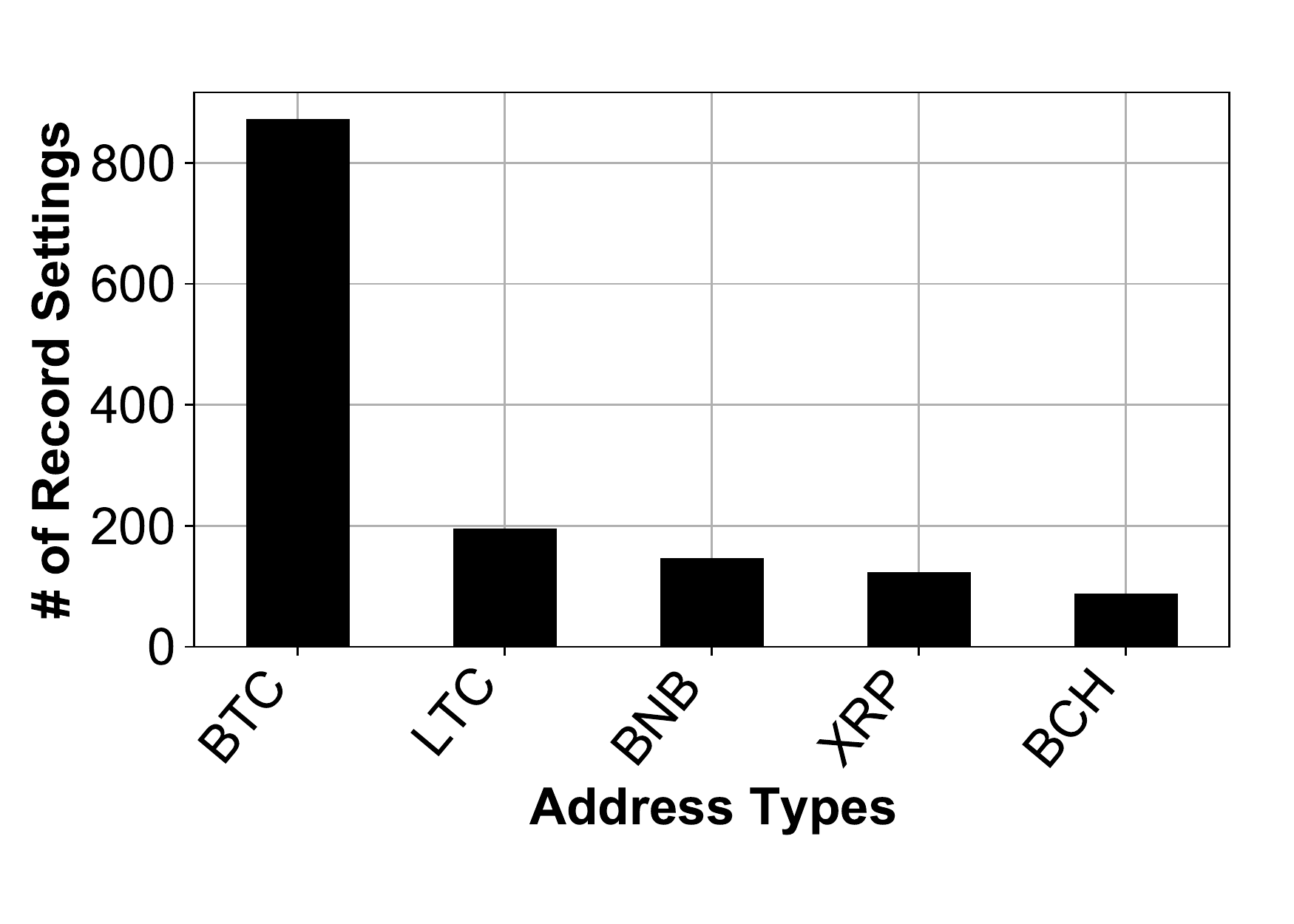}
\label{fig:ensaddr}
}
\hspace{0.05in}
\subfigure[The distribution of content hash records.]{
\includegraphics[width=0.22\linewidth]{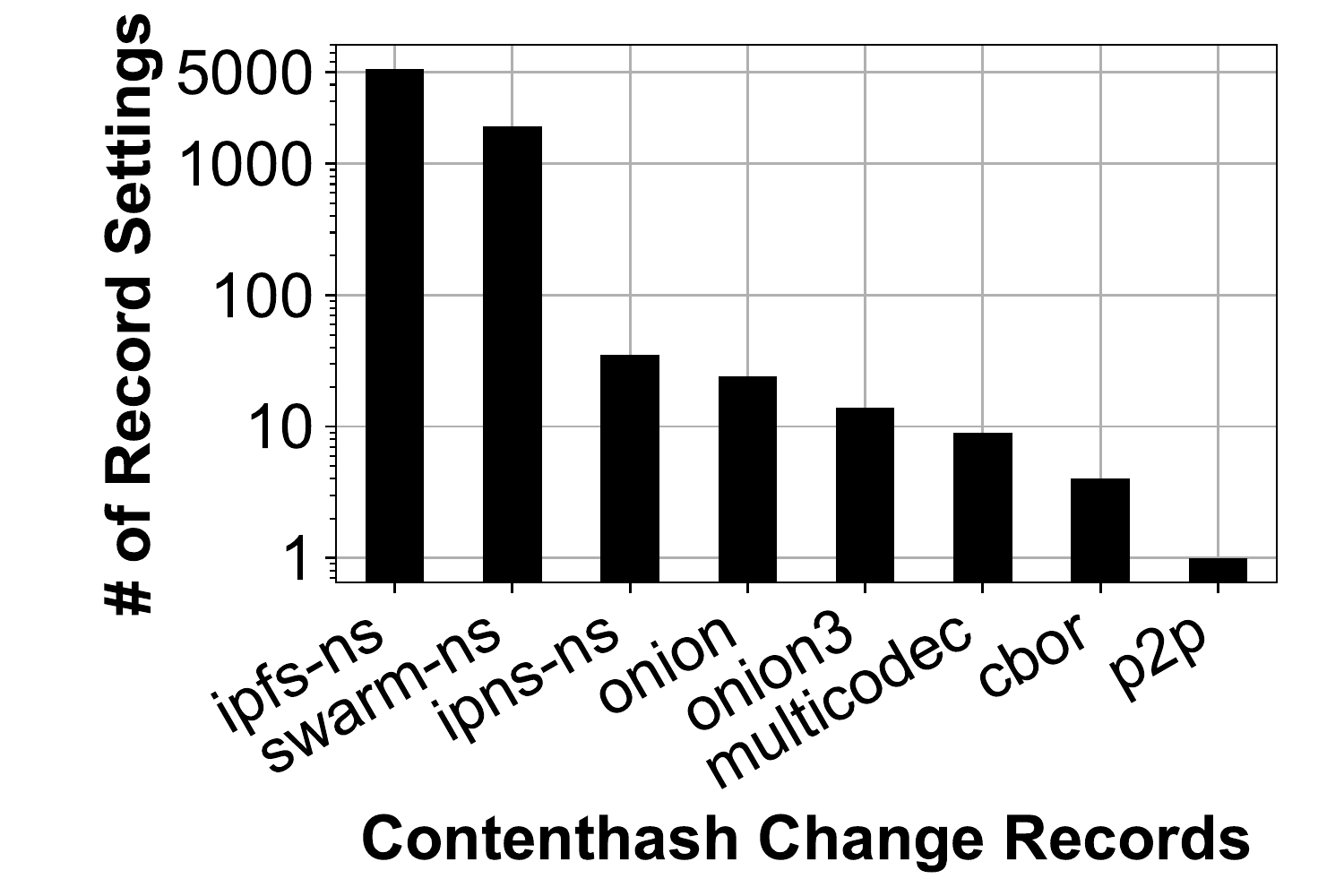}
\label{fig:contenthash}
}
\hspace{0.05in}
\subfigure[The distribution of top-9 text records.]{
\includegraphics[width=0.22\linewidth]{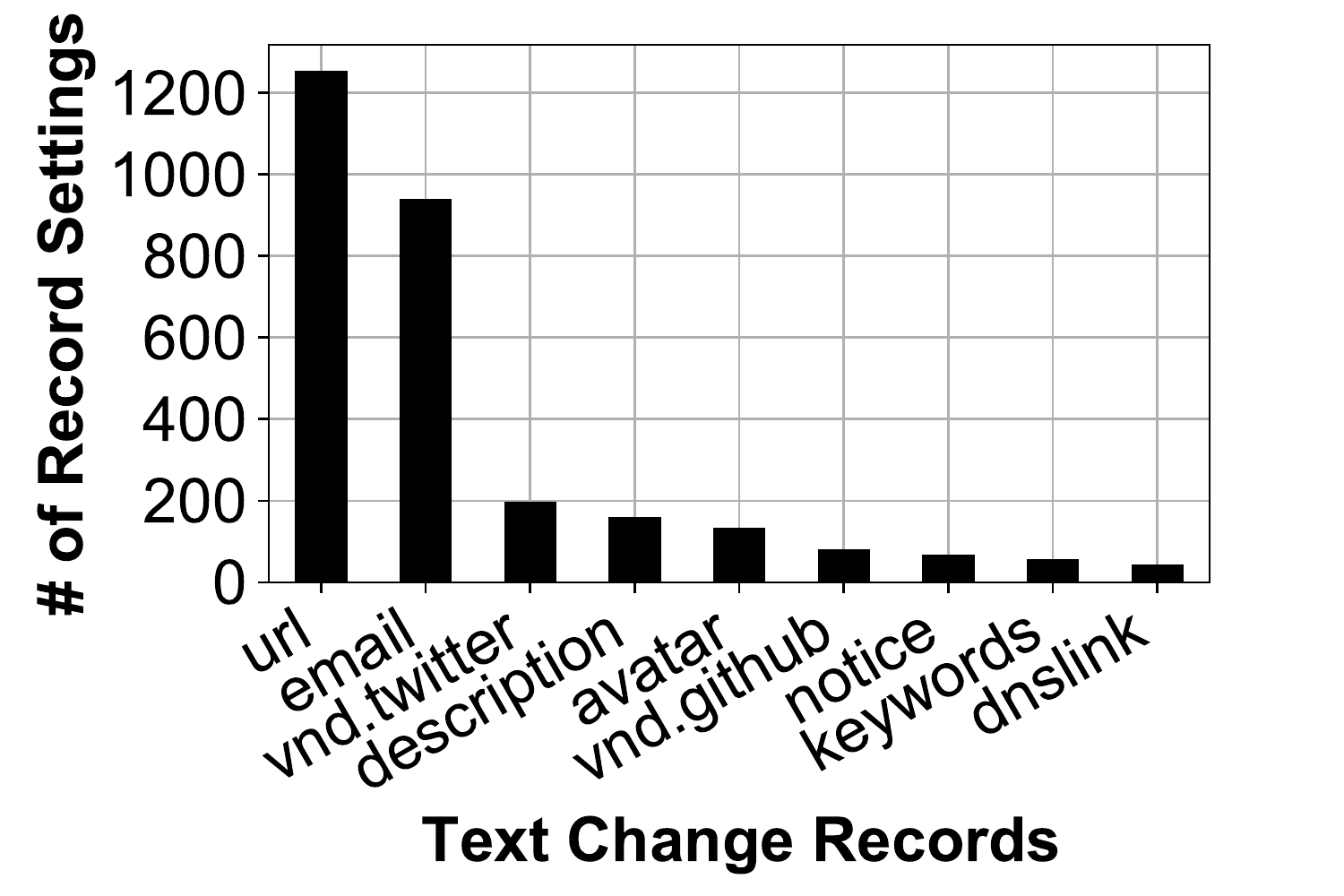}
\label{fig:textrec}
}
\hspace{0.05in}
\vspace{-0.2in}
\caption{The distribution of all records types and three major record types.
}
\end{figure}

Most of the address settings are relevant to Ethereum and BTC addresses. The former has $114,542$ setting records while the latter has 873 setting records. The distribution of top-5 non-ETH addresses is shown in Figure~\ref{fig:ensaddr}. Other types of addresses are set less than 200 times. It can be inferred that although addresses on other blockchains are supported, there are still few people taking advantage of ENS for purposes other than Ethereum addresses. 

\subsection{The use of content hash records} 
Another major use of ENS is to store content hashes, and roughly $5,300$ names have ever been set to content hash records. Thus, we further analyze how people use these hashes. There are roughly $2,700$ names with non-empty values and the distribution is shown in Figure~\ref{fig:contenthash}. Most of the content hashes (98\%) are set for IPFS and Swarm, two prominent solutions for decentralized publication and storage. Specifically, the ``ipns-ns'' hashes are used for InterPlanetary Name System (IPNS), which is designed for mutable contents~\cite{ipnsns}. Besides, a few names have been set for Tor \texttt{.onion} addresses while 10 of them have been set by ENS team~\cite{torsites}. For other records, our manual inspection suggests that they are all malformed IPFS hashes. For example, the eight ``multicodec'' hashes are generated by one user through the way of encoding IPFS hashes twice. The results show that ENS still has some room for dWeb resolutions. We will further investigate the malicious dWebs in Section~\ref{sec:malweb}. 

\subsection{The use of text records}
ENS allows users to set arbitrary text records in the form of key-value records with pre-defined keys at the same time. We perform an analysis on the keys of text records after excluding empty values, the top-9 of which are shown in Figure~\ref{fig:textrec}\footnote{The EIP-634 for text records updated on 2021 Jan. 27th, some of these keywords like ``vnd.twitter'' may be replaced by another keyword later.}. Most settings are for URLs, and we find that half of the history records are set to subdomains of OpenSea, suggesting that these names were for sale. Other URLs are used for official sites (e.g., \texttt{tokenfactory.global}), personal blog sites (e.g., \texttt{marvin-elsen.com}), etc. Besides predefined keywords like ``vnd.twitter'' (Twitter accounts) and ``description'' (description for names), etc., there are also some customized keywords in text records. For example, the ``dnslink'' records are mainly used by DAppNode~\cite{dappnode} for their dWebs and the ``gundb'' records are used for GunDB~\cite{gundb}, a P2P database. We identify 44 customized keywords in 214 record settings, indicating that people are exploring new ways to interacting with ENS. We will further analyze security issues related to URL records in Section~\ref{sec:malweb}.

\begin{framed}
\noindent \textbf{Answer to RQ2:} 
\textit{ENS is a fully open system where the ENS domain names can be set to any kinds of records. 
The most common use is for linking to blockchain addresses, which accounts for 67.3\% of record changes. Besides, it is also popular to use ENS names for dWebs and traditional websites, etc. We also find that people are exploring new ways to interact with ENS through text records. It suggests that ENS is on its way to becoming a complementary system of DNS.
}
\end{framed}

\section{Security Issues and Misbehaviors on ENS}
\label{sec:security}

In the following, we are committed to investigating whether there are any security issues on ENS, including both traditional issues and new issues introduced by the nature of ENS. 
For ENS names, we design an approach for searching and identifying possible \textit{squatting ENS names}, which is a common issue in traditional DNS (see Section~\ref{sec:squatting}). 
For ENS name records, we are interested in whether there are \textit{malicious addresses or websites} set in ENS names, which corresponds to DNS malicious websites (see Section~\ref{sec:malweb} and Section~\ref{subsec:scamaddress}). In particular, since ENS supports resolutions to dWebs that cannot be taken down easily, we also resort to ENS content hash records to find if there are malicious dWebs. Besides, since DNS and ENS have different designs, we take efforts to check \textit{whether the design of ENS would increase the attack surface} (see Section~\ref{subsec:recordattack}).

\subsection{Domain Squatting}
\label{sec:squatting}
Domain squatting is an act of registering domains with the same or confusingly similar names with famous landmarks, which is usually used for malicious intents. Domain squatting issues have been extensively studied in traditional DNS.
To understand whether this common malicious act is prevalent in ENS, we analyze the registered ENS domain names from three perspectives.
First, different with the domain squatting on DNS, where famous brand names with most TLDs are registered by brand owners, the \texttt{.eth} TLD introduced by ENS is a new namespace. As mentioned in Section~\ref{sec:phase2}, there are few brand claiming their \texttt{.eth} names, which leaves room for explicit domain squatting. 
Second, as observed in previous research on DNS~\cite{typosquatting}, attackers tend to use domain typo-squatting methods to register domain names that are similar to well-known DNS domains. Thus, the same methods could be exploited to register ENS names.
At last, since ENS introduced new \texttt{.eth} TLDs, squatters are more likely to register more ENS names besides the top domain names in the Alexa list. It is interesting to investigate how many squatters have taken such actions. 

\subsubsection{Explicit Squatting of Known Brands}  
\label{sec:popreg}
As stated in Section~\ref{sec:studydesign}, we have taken advantage of Alexa top-100K name list to match each level of ENS names. In other words, we calculate the labelhash (keccak256 hash of the name, see Section~\ref{sec:ensback}) of each 2LDs in Alexa list and match them with each labelhash in ENS names\footnote{To remove possible false positives, we exclude $119,764$ Alexa names and $15,701$ typo-squatting variants that have a length of less than 4.}.
Among top-100K names in Alexa, there are $18,233$ top Alexa names that could be found in ENS native 2LDs, i.e., we find that $18,233$ ENS \texttt{.eth} names (only 31 of which were claimed in short name claim period) have the same names with names of Alexa list in total. Note that, not all matching ENS names are squatting names since the name could be registered by the actual owner. 
Here, our heuristic is that, if one Ethereum address owns more than one known ENS names (e.g., \texttt{google.eth} and \texttt{facebook.eth}) that belong to different owners in their DNS domain names, we consider these \texttt{.eth} names of known brands are explicit squatting names. 
For example, the address \texttt{0x782cf6b6e735496f7e608489b0c57ee27f407e7d} have registered \texttt{google.eth}, \texttt{mcdonalds.eth}, \texttt{redbull.eth}, etc., and these brands are not belonging to the same owner. Thus, we believe this address is involved in the explicit ENS name squatting.

Through this way, $15,179$ ENS \texttt{.eth} squatting names controlled by $1,532$ Ethereum addresses are found. Top-10 holders are shown in Table~\ref{tab:topalexaname} of Appendix. Many famous brands were registered for squatting. Among these explicit squatting ENS \texttt{.eth} names, over $42.7$\% of \texttt{.eth} names are active now, suggesting that some attackers tend to hold these explicit squatting names for a long time.

\subsubsection{Typo-Squatting ENS Names} 
\label{sec:typosqu}
To detect typo-squatting ENS names, we take advantage of dnstwist~\cite{dnstwist}, a widely used tool to generate typo-squatting variants of Alexa top-100K names. Dnstwist can generate $12$ kinds of squatting variants. Taking domain \texttt{google.com} as an example, dnstwist can generate $1,982$ variants on this domain through different methods, such as addition (e.g., \texttt{googlea.com}), bitsquatting (e.g., \texttt{ggogle.com}), homoglyph (e.g., \texttt{googlë.com}), hyphenation (e.g., \texttt{g-oogle.com}
), insertion (e.g., \texttt{g0oogle.com}), omission (e.g., \texttt{googe.com}), repetition (e.g., \texttt{googgle.com}), replacement (e.g., \texttt{googl4.com}), transposition (e.g., \texttt{goolge.com}), vowel-swap (e.g., \texttt{geogle.com}), and various (e.g., \texttt{googlecom.com}). We feed all the domains in Alexa top-100K to dnstwist and get $755,908,096$ variants. Similarly, we calculate the labelhash of their 2LDs to check whether these squatting names have already been registered on ENS. 

As a result, we have identified $18,483$ ENS typo-squatting \texttt{.eth} names targeting $13,450$ Alexa domain names, examples of which are shown in Table~\ref{tab:toptyponame} of Appendix. 
Figure~\ref{fig:namevariants} shows the distribution of variant types. There are roughly 5K bitsquatting variants and $1,270$ homoglyph domains. 
Over 52\% of the typo-squatting ENS \texttt{.eth} names are active by the time of our study. 

\begin{figure}[t]
\begin{tabular}{cc}
\begin{minipage}[t]{0.48\linewidth}
    \includegraphics[width=1\textwidth]{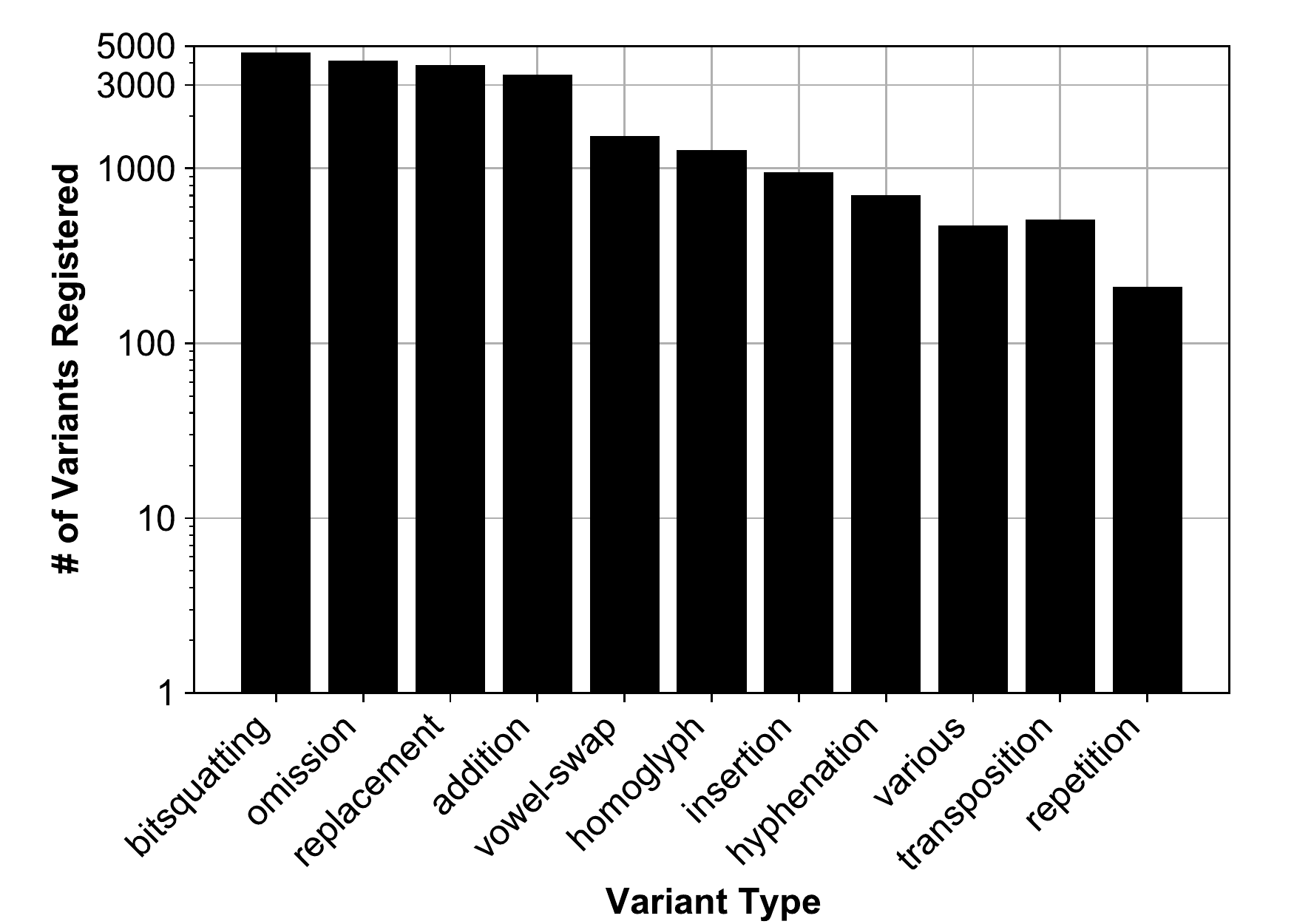}
    \caption{The distribution of squatting variant types.}
    \label{fig:namevariants}
\end{minipage}
\hspace{0.1in}
\begin{minipage}[t]{0.48\linewidth}
    \includegraphics[width=1\textwidth]{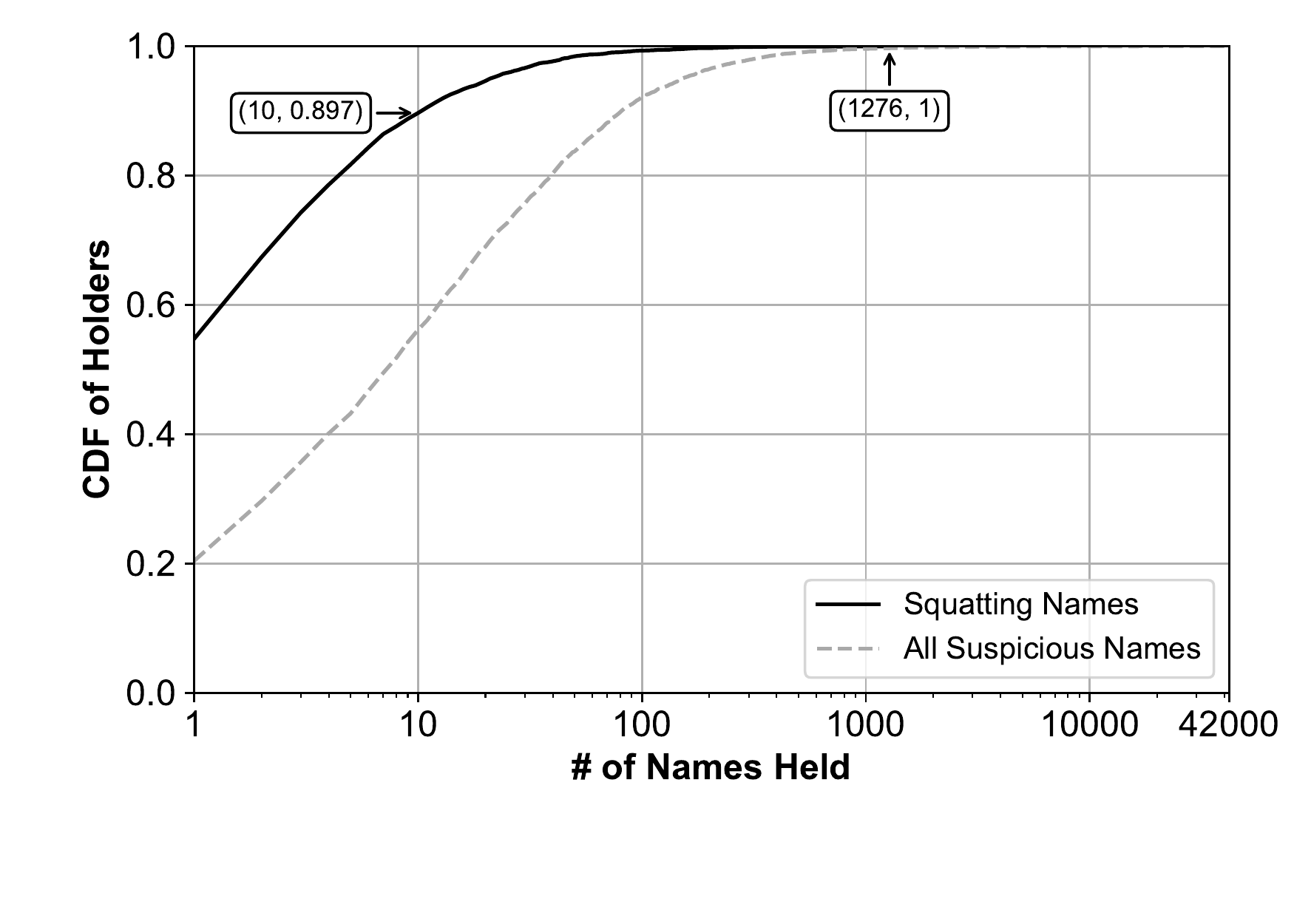}
    \caption{The ENS squatting/suspicious name distribution of ENS name holders.}
    \label{fig:susnamehold}
\end{minipage}
\end{tabular}
\vspace{-0.1in}
\end{figure}

\subsubsection{Squatting Names Analysis}

In total, we collect $33,662$ unique squatting ENS \texttt{.eth} names based on previous heuristics. We next investigate the records and holders of these squatting domains, seeking to identify more suspicious names based on ``guilt-by-association'' expansion
method. 

\textbf{The records of squatting names.} 
Only $4,474$ squatting ENS \texttt{.eth} names ($3,775$ active ones) have been set records and most (85\%) of them were set only blockchain address records. 
Besides Ethereum addresses, most of their other records are related to sales like Opensea links, IPFS websites posting sale information, etc. It indicates the squatting nature of these identified names.

\begin{figure}[h]
\centering
\includegraphics[width=0.7\textwidth]{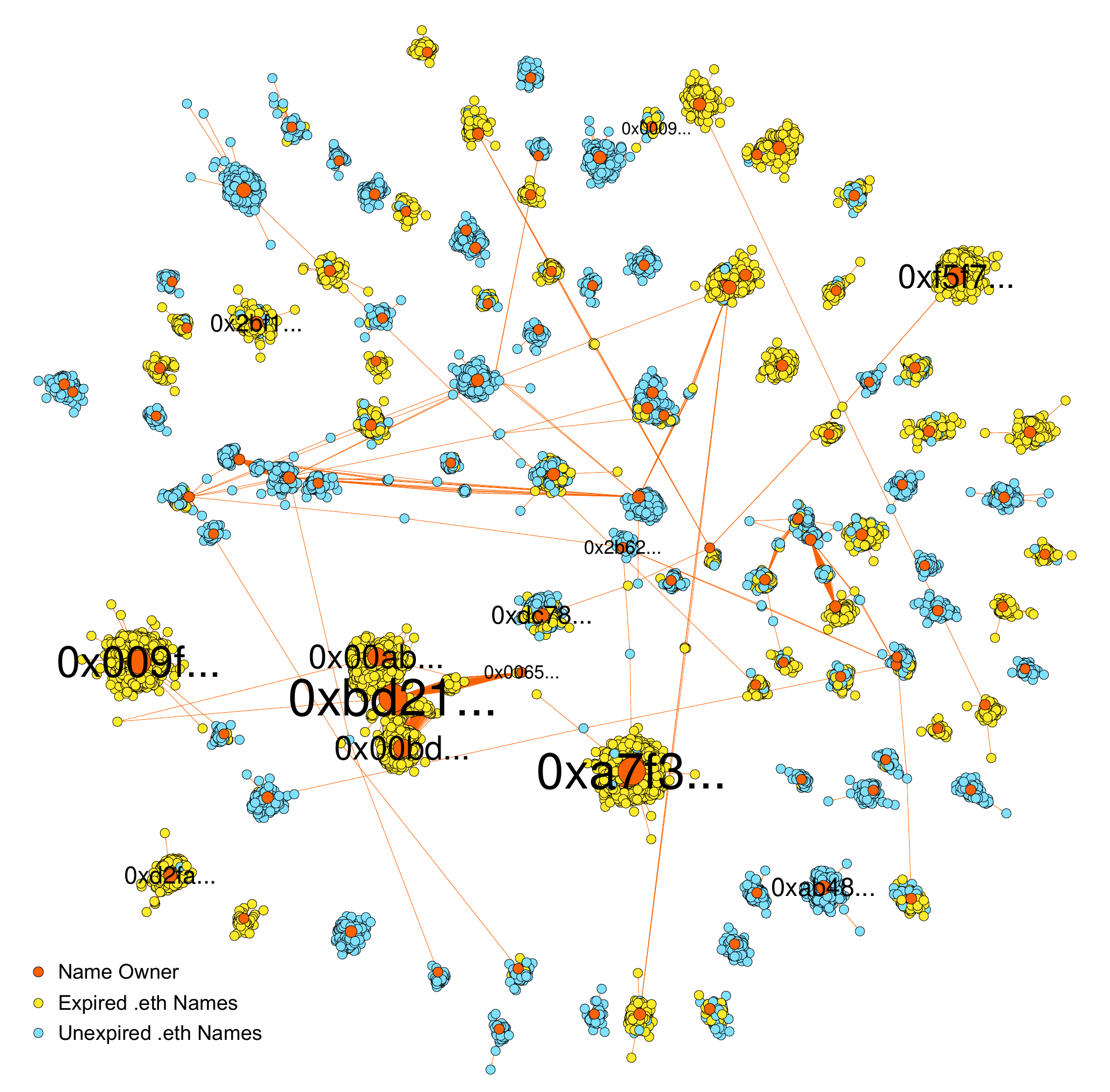}
\vspace{-0.1in}
\caption{The distribution of all time ownership of squatting names. Note that we exclude addresses that ever have less than 50 ENS names and their corresponding names for showing the figure clearly.}
\vspace{-0.2in}
\label{fig:squensholder}
\end{figure}

\textbf{The relations between addresses.}
For the identified squatting ENS \texttt{.eth} names, we further investigate the \textit{name-to-holder} and \textit{holder-to-holder} relations. Overall, these $33,662$ names have been ever owned by $6,548$ addresses. Figure~\ref{fig:squensholder} shows the relations between names and owners according to the all-time ownership (note that an ENS domain name can change its ownership). It is interesting to see that, some names have ever been owned by more than one address. Further investigation reveals that some addresses also transferred their names. For example, the address \texttt{0xbd21109e2bdcb24c4fbcdc16a4c90f34e81228e2} received ENS \texttt{.eth} names from 3 addresses in the figure. It can be also seen that several addresses are holding a large number of ENS \texttt{.eth} names. As shown in Figure~\ref{fig:susnamehold}, over 10\% of addresses hold more than 10 squatting ENS \texttt{.eth} names, and these names account for 70\% of all squatting ENS names in total.

\textbf{``Guilt-by-association'' Expansion.} 
Our heuristic is that, for the squatters that register squatting domains, it is quite possible that they would hold more squatting domains than we identified, as we only consider the most popular Alexa domains in our previous exploration. We call it the ``Guilt-by-association'' principle, which has been used in previous work to identify malicious domains and malware~\cite{sebastian2020towards,khalil2018domain}. Thus, we analyze all the ENS names held by the identified squatters. In total, we find $279,193$ suspicious squatting \texttt{.eth} names, which we believe are high suspicious to be squatting names too. 
Figure~\ref{fig:susnamehold} shows the distribution of these addresses. Over 40\% of the squatters have ever held more than 10 ENS \texttt{.eth} names, which account for over 96\% of all suspicious names in total. The top-10 holders that have most squatting names are shown in Table~\ref{tab:topsushold} of Appendix. The top address \texttt{0xbd21109e2bdcb24c4fbcdc16a4c90f34e81228e2} has ever acquired $1,276$ ENS squatting names, and it has ever held more than 40K names in total. Most of the ENS names owned by it contain Chinese pinyin (e.g., \texttt{jianshu.eth}) or numbers (e.g., \texttt{8062222.eth}), which were transferred from 6 other addresses including the 4 addresses mentioned in Section~\ref{sec:general}. These top-10 addresses have ever held around 17\% of all names, which is not a number that could be ignored. These observations show that squatters are likely to register more other ENS names even when these names don't rank top in Alexa list.

\begin{figure}[h]
\centering
\includegraphics[width=0.7\textwidth]{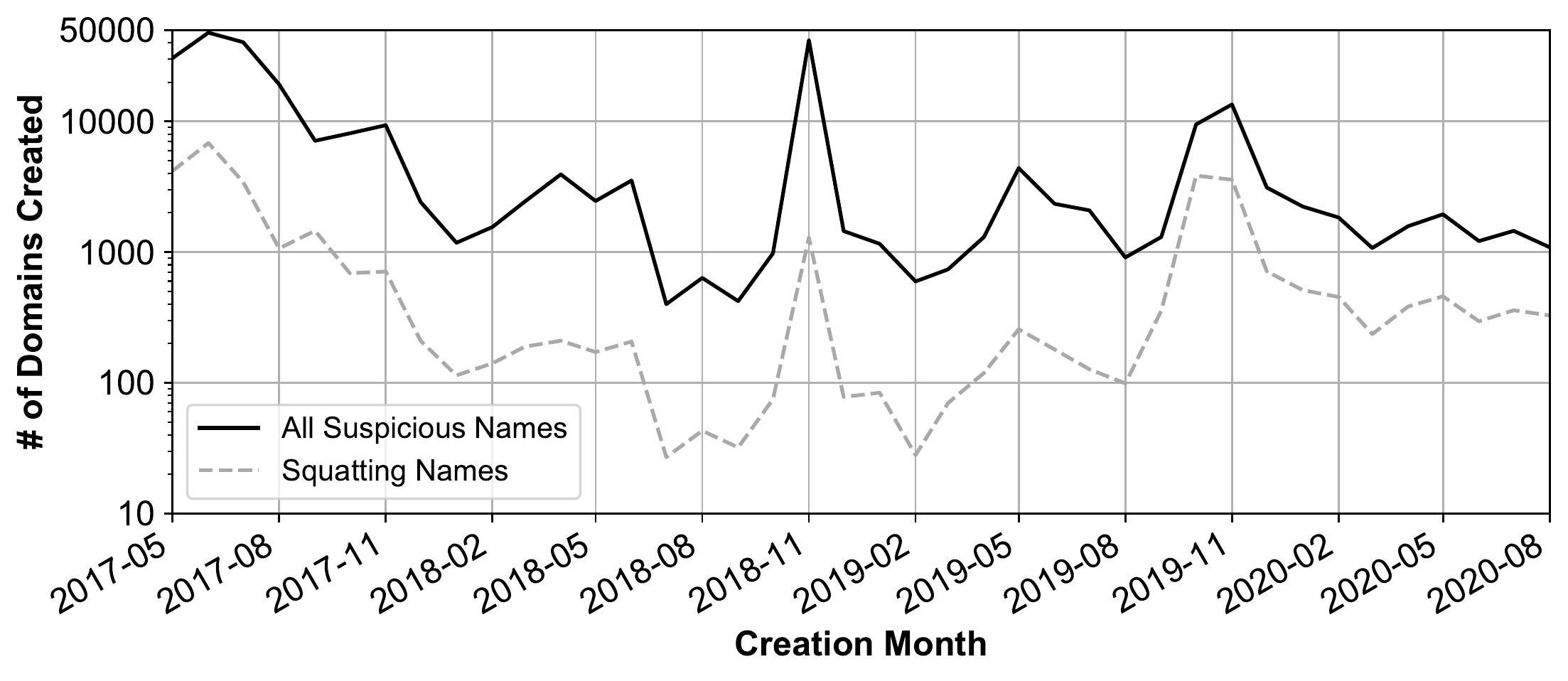}
\vspace{-0.1in}
\caption{The evolution of squatting names we identified.}
\vspace{-0.1in}
\label{fig:squnameevo}
\end{figure}

\textbf{The evolution of squatting names.}
Further, we also investigate the evolution of all these suspicious ENS squatting names, as shown in Figure~\ref{fig:squnameevo}. The first batch of squatting names (e.g., \texttt{zhifubao.eth}, pinyin of Alipay) were registered around 2017 May 9th, almost at the same time as the initial auction start. Furthermore, the overall squatting trend follows the trend of general names we study in Section~\ref{sec:general}, suggesting to some extent that there have been certain squatting behaviors in each period of ENS. However, after ENS team launched the permanent registrar, most expired names were given up by these squatters but  $67,614$ suspicious ENS squatting names (75\% of all active ENS \texttt{.eth} names by the time of the study) are still held. For example, when the massive squatting behaviour happened on 2018 Nov., the address \texttt{0xbd21109e2bdcb24c4fbcdc16a4c90f34e81228e2} registered more than 40K names while by the study time he owns only one name.

\subsection{Websites with Misbehaviors}
\label{sec:malweb}

Since ENS provides the functionality to store URL information in ``TextRecord" or ``DNSRecord'', and store decentralized web identifiers or onion website hashes in ``ContentHash'', there is a possibility that some bad actors can exploit these functions to deliver malicious or illegal web contents. Thus we attempt to perform an analysis on websites that have ever been stored in ENS. As mentioned in Section~\ref{sec:records}, we get 5,879 unique dWeb hashes, 34 onion hashes and 620 URLs.

\subsubsection{Method.}
We first upload all the URLs to VirusTotal~\cite{virustotal}, which is a famous anti-virus service that providing over 70 anti-virus engines, for scanning malicious websites. Following previous studies~\cite{peng2019opening,he2020virus}, if a URL is reported by 2 or more anti-virus engines, it will be marked as a malicious URL. 
Second, we use Eyewitness~\cite{eyewitness} to get the screenshots and source codes of all these websites, which will be uploaded to Google Cloud Natural Language API and Vision API~\cite{googlevision} to check whether they contain censored (e.g., adult and gambling) contents. Besides, we also mark the URLs that have keywords like ``casino'' or ``generator'' in their names or contents as suspicious URLs. All the suspicious URLs will be manually inspected to reduce false positives. 

\subsubsection{Result.} 
In total, we get 19 (17 second-level ENS names) malicious dWeb URLs. Examples are shown in Figure~\ref{fig:scamdweb} of Appendix. The malicious websites we find are involving in gambling (7), adult (5) or scam activities (7). Note that, since dWebs may not store the data online persistently, some content cannot be reached during the analysis period. Thus, the actual number of malicious dWebs should be higher than we identified. Nevertheless, due to the decentralized nature, these malicious dWebs have the ability to exist online for a long time, causing certain losses to users. For traditional URLs in text records, although we do not find any malicious traditional DNS websites on ENS and there is currently no convenient way to access DNS domains through ENS, there may be a risk of malicious use of this feature in the future.

\subsection{Scam Address}
\label{subsec:scamaddress}
As mentioned in Section~\ref{sec:records}, ENS is used mainly for storing Blockchain address records. Thus, we further seek to analyse whether any addresses are used for malicious purpose.

\subsubsection{Method.}
Considering that there is no available comprehensive dataset of scam blockchain addresses, we first compile a scam address list from various sources. Etherscan and Bloxy~\cite{bloxy} have labelled a list of ``phishing'' or ``hacked'' Ethereum addresses. BitcoinAbuse~\cite{bitcoinabuse} and CryptoScamDB~\cite{cryptoscamdb} are hosting websites for tracking malicious blockchain addresses. We crawl all the addresses above and get over 58K scam addresses in total. 
Then, we match the addresses stored in ENS with the scam address list. 

\subsubsection{Result.}
We find three scam addresses that have been registered in the ENS, which is shown in Table~\ref{tab:maladdr}. Figure~\ref{fig:airdrop} in Appendix shows the screenshot of an airdrop scam using ENS name \texttt{valus.smartaddress.eth} to promote their scam token. Our manual inspection suggests that the second address may not be a scam address, but its ENS name was linked to a Ponzi-like website. 
Despite few occurrences, malicious actors are finding a way to exploit ENS in malicious activities.

\begin{table*}[]
\small
\centering
\caption{Identified suspicious scam addresses in ENS.}
\vspace{-0.1in}
\resizebox{0.9\linewidth}{!}{

\begin{tabular}{@{}lll@{}}
\toprule
ENS names              & Address                                         & Description  \\ \midrule
\texttt{valus.smartaddress.eth} & ETH: 0x903bb9cd3a276d8f18fa6efed49b9bc52ccf06e5 & An airdrop scam \\ \midrule
\texttt{four7coin.eth} &
  BTC: 385cR5DM96n1HvBDMzLHPYcw89fZAXULJP &
  \begin{tabular}[c]{@{}l@{}}Reported as a Ponzi scheme by   BitcoinAbuse,\\ actually is a Bittrex cold wallet~\cite{bittrex}\end{tabular} \\ \midrule
\begin{tabular}[c]{@{}l@{}}\texttt{jessica.chainlinknode.eth},   \\ \texttt{jessica.atethereum.eth}, \\ \texttt{crunk.eth}\end{tabular} &
  BTC: 1F1tAaz5x1HUXrCNLbtMDqcw6o5GNn4xqX &
  \begin{tabular}[c]{@{}l@{}}Reported to be ransomware address,\\ actually is a old Silkroad seized wallet~\cite{silkroad}\end{tabular} \\ \bottomrule
\end{tabular}
}
\vspace{-0.1in}
\label{tab:maladdr}
\end{table*}

\subsection{Record Persistence}
\label{subsec:recordattack}

\subsubsection{Attack Scenario} 
\begin{figure}[t]
\begin{tabular}{lr}
\begin{minipage}[t]{0.42\linewidth}
    \includegraphics[width=1\textwidth]{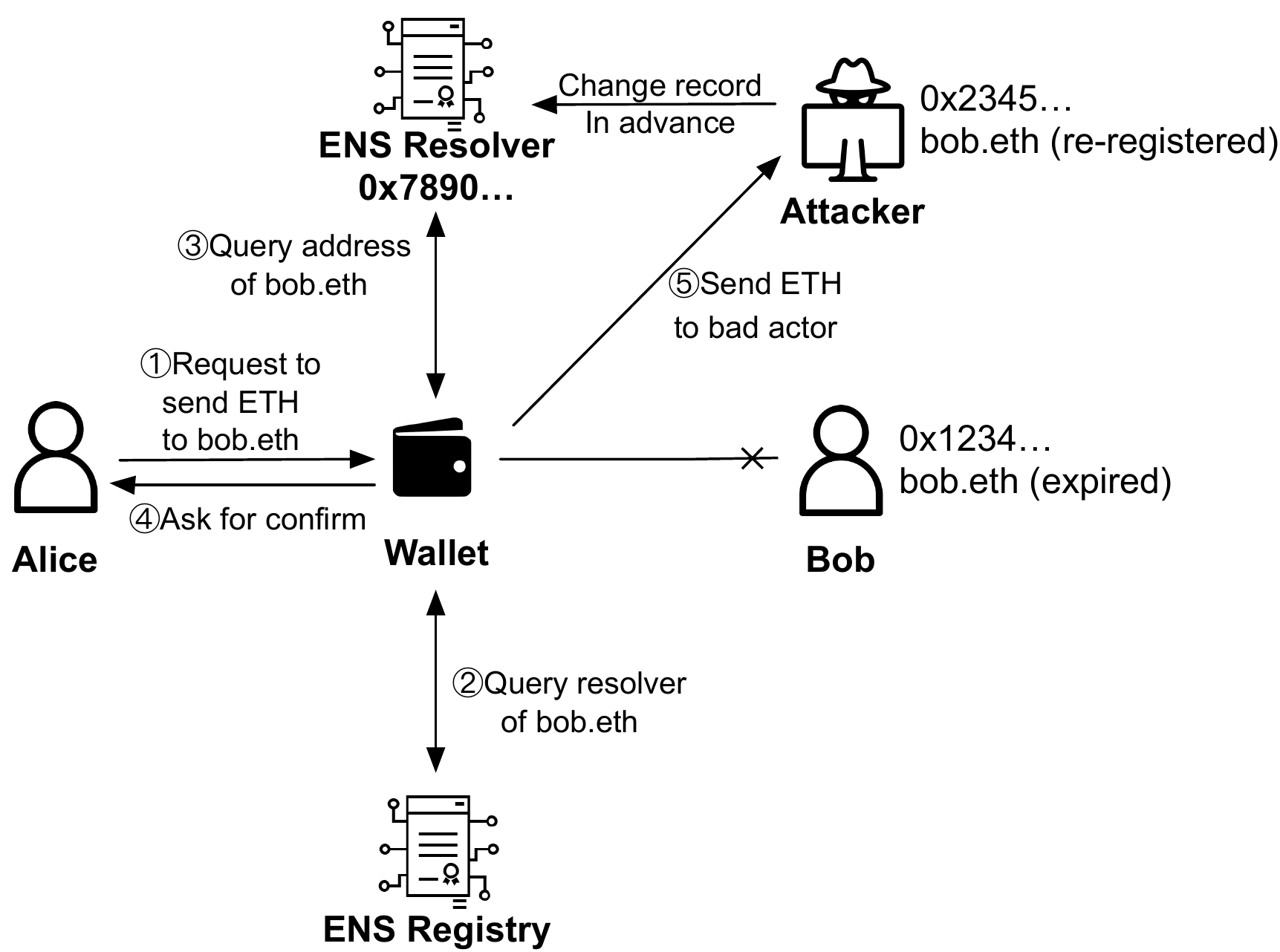}
    \caption{An Example of the attack scenario that exploits the record persistence issue.}
    \label{fig:recordissue}
\end{minipage}
\hspace{0.2in}
\begin{minipage}[t]{0.42\linewidth}
    \includegraphics[width=1\textwidth]{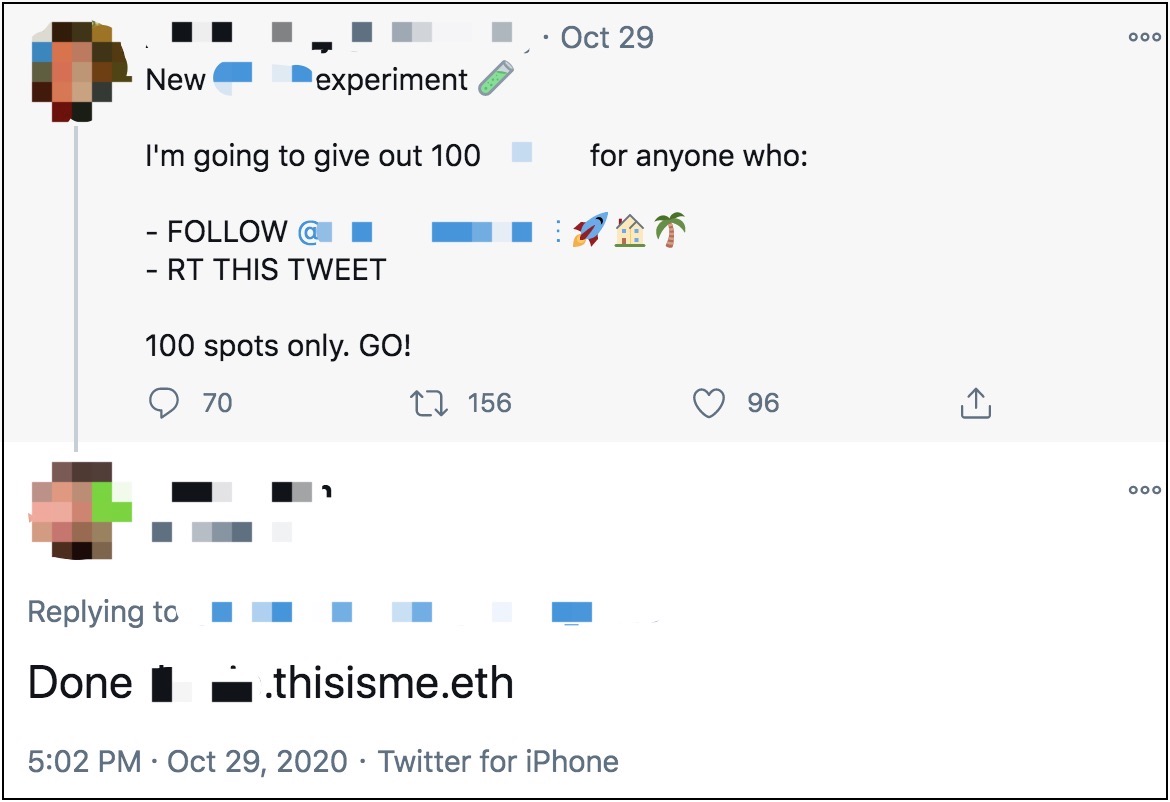}
    \caption{Some people are still using \texttt{thisisme.eth} names.}
    \label{fig:thisisme}
\end{minipage}
\end{tabular}
\vspace{-0.2in}
\end{figure}

We find that when an ENS name expires, the name and its subdomain names' records would be kept and some ENS-supported wallets could still resolve them to blockchain addresses or other records. 
This may lead to scams when previous users are still using these expired ENS names. An example of the possible attack is shown in Figure~\ref{fig:recordissue}. When an attacker re-registers the expired name that has records and changes the name's records to his address in advance, people who are unaware of this change and do not check the recipient addresses (this is originally one of the purposes using ENS) would eventually send money to the attacker.

\subsubsection{Vulnerable ENS Names.}
We further make an attempt to find how many ENS names are vulnerable to such kind of attacks, i.e., the expired ENS names that have records.
After resorting to history records, we find that $16,017$ expired \texttt{.eth} names have records within them or their subdomains ($3,116$ subdomains), examples of which are shown in Table~\ref{tab:expsubname} of Appendix. 
These names contain not only blockchain addresses but also dWeb hashes that may direct users to phishing/malicious websites if abused. In particular, the ENS name \texttt{thisisme.eth} has 706 subdomain names and all of them have Ethereum address records. It was listed in \texttt{enslisting.com} for free registration and was transferred to a smart contract to ensure that subdomain name records of this name would not be modified easily. Also, the ENSListing team claimed that they would cover the cost of annual renewals~\cite{enslisting}. 
However, the name was expired on May 4th 2020, and we re-registered it for protection. 
Note that there are still some people using subdomain names of this name. 
Figure~\ref{fig:thisisme} shows an example, where a man was taking part in an airdrop activity with his \texttt{thisisme.eth} names. 
As we mentioned, attackers can identify the vulnerable ENS names and re-register them for making a profit. Although ENS team has proposed a new function of email notifications to remind people to renew their names on June 2020~\cite{email}, it is still a severe security concern of ENS names.

\begin{framed}
\noindent \textbf{Answer to RQ3:} 
\textit{
Although the working mechanism of ENS has been improved greatly during the evolution, there still leaves room for attackers to abuse the system. We have identified several security issues and misbehaviors including traditional DNS security issues and new issues introduced by ENS smart contracts. A few squatters are found holding a lot of famous brand names and their variants, which could be used for malicious purposes. Some malicious decentralized websites and scam addresses are found in records of ENS names. Besides, a record persistence issue is found that may cause potential attacks, which could lead to financial loss to users.
}
\end{framed}
\section{Discussion}

\textbf{Implications}
Our findings are very important to the stakeholders in the ecosystem. Considering the certain number of active addresses and users on ENS, along with the integrated dApps and DNS TLDs, ENS still has a relatively healthy ecosystem. According to the official team~\cite{ensdnstest,ens2021}, they have deployed a smart contract on Ropsten testnet to enable the integration of most DNS 2LDs and planned to make many improvements including scaling on Layer 2~\cite{layer2}, etc., which will greatly facilitate the daily use and reduce the cost of ENS users.
However, since we find several security issues on ENS, the ENS team may consider solutions to enhance the security of ENS. Considering the difficulty and cost of making changes to the deployed smart contracts, developers of dAPP or blockchain wallets that support ENS may deploy similar methods used in this work to detect squatting names or malicious records and give reminders to users who are trying to interact with these suspicious ENS names. For ENS users, it could be better to check the real addresses under ENS names when interacting with them.
Besides, as BNS shares some common properties like tamper-resistant records, methods used in this work could be extended to study other BNS systems.

\noindent \textbf{Limitations}
First, we have only restored 86.6\% of all \texttt{.eth} names to their readable names (see \S\ref{sec:dataset}), which could bring limitations when we identify squatting names since we may not detect some combo-squatting~\cite{combosquatting} ENS names. Nevertheless, our dataset of recovered names is the largest dataset, which has no effect on our study of explicit name squatting and typo-squatting since we could calculate their hash values.
Besides, there are many cyber attacks targeting at DNS, while some attacks like DDoS are not studied due to time or cost, which could be explored in future work.

\section{Related Work}

\textbf{Design of BNS Systems.}
Many researchers have been studying the design of blockchain-based DNSs. Hari et al.'s work~\cite{hari2016internet} is one of the first works to propose the design of blockchain-based DNS. They analyzed the limitations of the traditional DNS and their dependencies on Public Key Infrastructures (PKIs). Then, they proposed a distributed, tamper-resistant DNS infrastructure. Similarly, Guan et al.~\cite{guan2019authledger} presented a domain authentication scheme named AuthLedger to reduce the level of trust in certificate authorities (CAs). 
On the other hand, some studies are focused on improving the security of the DNS nodes~\cite{he2020td,liu2018data,wang2019blockzone,zhang2019research}. For example, He et al.~\cite{he2020td} proposed a trustworthy decentralised DNS root management architecture based on permissioned blockchain. Besides, Gourley et al.~\cite{gourley2018blockchain} proposed an improved DNSSEC based on blockchain, which provides the same security benefits as DNSSEC whilst addressing its drawbacks. 

\noindent \textbf{Analysis of BNS Systems.}
A few studies have analyzed different kinds of blockchain-based DNS systems~\cite{al2020brief,ali2020improving,wei2017review,bagay2020blockchain,patsakis2020unravelling,karaarslan2018blockchain,liu2019comparative,kalodner2015empirical}. Kalonder et al.~\cite{kalodner2015empirical} firstly performed a empirical analysis on Namecoin. 
After that, some other studies have characterized mainly on the properties of different kinds of blockchain-based DNS. Patsakis et al.~\cite{patsakis2020unravelling} surveyed the threat of blockchain-based DNS including malware, underlying registrar mechanism, domain market, phishing, motivation and immutability. 
Very few studies have mentioned ENS. For example, Liu et al.~\cite{liu2019comparative} and Karaarslan et al.~\cite{karaarslan2018blockchain} compared the designs of several blockchain-based DNSs including ENS. However, there lacks a systematic study of ENS, including the status quo of ENS, security issues, etc.

\section{Conclusion}
We take the first step to systematically characterize Ethereum Name Service. By collecting and analyzing large-scale ENS event logs, we obtain a number of interesting observations and reveal security issues related to ENS. We believe ENS is a promising system on its way to being complementary to DNS, but the spams and security issues may impede its progress. Our efforts in this paper can positively contribute to the BNS ecosystem and offer practical insights. 

\balance

\bibliographystyle{plain}
\bibliography{cite}

\appendix
\renewcommand{\thesection}{Appendix~\arabic{section}}
\setcounter{section}{0}

\section{Appendix for study design in Section~\ref{sec:studydesign}}
\subsection{The additional resolvers.}
Table~\ref{tab:exresolver} shows eight open-source additional resolvers we add, the names of which were fetched from Etherscan labels and search engines.
\begin{table}[h]
\small
\centering
\caption{The eight additional resolvers we add.}
\begin{tabular}{@{}ccc@{}}
\toprule
Resolver Name             & Address                                    & \# of Event Logs \\ \midrule
ArgentENSResolver1        & 0xDa1756Bb923Af5d1a05E277CB1E54f1D0A127890 & 67,068           \\
OldPublicResolver3        & 0x5FfC014343cd971B7eb70732021E26C35B744cc4 & 27,693           \\
OldPublicResolver4        & 0xD3ddcCDD3b25A8a7423B5bEe360a42146eb4Baf3 & 6,610            \\
AuthereumEnsResolverProxy & 0x4DA86a24e30a188608E1364A2D262166a87fCB7C & 8,406            \\
OpenSeaENSResolver        & 0x9C4e9CCE4780062942a7fe34FA2Fa7316c872956 & 226              \\
ArgentENSResolver2        & 0xb23267e7a0DEe4DCBA80C1D2FFDb0270aF76fe80 & 478              \\
PortalPublicResolver      & 0x0B3eBEccC00E9CEae2BF3235d558EdA7398BE91E & 236              \\
TokenResolver             & 0x074d58C0a0903d4C7DB9388205232602a0bF9Bf0 & 152              \\ \bottomrule
\end{tabular}
\label{tab:exresolver}
\end{table}

\section{Appendix for general overview in Section~\ref{sec:general}}
\subsection{Top-10 .eth name owners by study time.}
Table~\ref{tab:top10owner} shows top-10 owners that hold most of \texttt{.eth} names by the time of the study.

\begin{table}[h]
\caption{The top-10 \texttt{.eth} name owners by the time of this study.}
\begin{tabular}{@{}ccc@{}}
\toprule
Owner Address & \begin{tabular}[c]{@{}c@{}}.eth Names Owned \\ (By Study Time)\end{tabular} & \begin{tabular}[c]{@{}c@{}}.eth Names Owned \\ (All Time)\end{tabular} \\ \midrule
0xbcbd4885ee8b2b74249c5ad9b8b668fb256a51b1 & 2262 & 2367 \\
0xea32bf2135888c46157320f9fe3539211945cbae & 1388 & 1388 \\
0x8124b02b4967bd7338a450420a572d574915ff9c & 916  & 1040 \\
0xe2faa63f2351c6f2b88659f2fffc2167172d329a & 806  & 906  \\
0xddd3964d75d59b6b6d5c31eb313bba5ebf076364 & 704  & 757  \\
0xab48edd90bdf367d326d827758bacd2460c59d17 & 655  & 657  \\
0x1d6f2f0356b3defadf14b1a0f8a3dcda89367d68 & 634  & 642  \\
0xf14955b6f701a4bfd422dcc324cf1f4b5a466265 & 567  & 766  \\
0xdc784458e2516a9d0531509f784dcf21983a60d7 & 543  & 1308 \\
0x813ec5facf289bd41365a8f1c9038a1228e95201 & 530  & 542  \\ \bottomrule
\end{tabular}
\label{tab:top10owner}
\end{table}

\subsection{Top-10 valuable ENS names.}
During the Vickrey auction period, the top-10 valuable ENS names are shown in Table~\ref{tab:top10auction}.

\begin{table}[h]
\small
\centering
\caption{Top-10 valuable ENS names.}
\resizebox{0.8\linewidth}{!}{
\begin{tabular}{@{}crcc@{}}
\toprule
Name & \multicolumn{1}{c}{\begin{tabular}[c]{@{}c@{}}Auction \\ Price\\ (ETH)\end{tabular}} & \begin{tabular}[c]{@{}c@{}}\# of \\ Valid \\ Bids\end{tabular} & Owner \\ \midrule
darkmarket.eth & 20,103 & 4  & 0x8759b0b1d9cba80e3836228dfb982abaa2c48b97 \\
openmarket.eth & 10,055 & 10 & 0x8759b0b1d9cba80e3836228dfb982abaa2c48b97 \\
exchange.eth   & 6,660  & 72 & 0xdc9fbececa49457fbcb2ee1dfd576a8be06a5c30 \\
blackjack.eth  & 5,910  & 48 & 0x217ad132e6271a9a19a818bde4cb6498888a3ba2 \\
tickets.eth    & 2,977  & 39 & 0x8759b0b1d9cba80e3836228dfb982abaa2c48b97 \\
payment.eth    & 2,600  & 66 & 0x8759b0b1d9cba80e3836228dfb982abaa2c48b97 \\
trading.eth    & 2,563  & 37 & 0x752975f5990c33da38c4cd50f0a41b70b3a6796c \\
registry.eth   & 2,509  & 12 & 0x352f25babf4a690673e35195efa8f79d05848aad \\
jackpot.eth    & 2,123  & 21 & 0x352f25babf4a690673e35195efa8f79d05848aad \\
lottery.eth    & 1,700  & 26 & 0x9a02ed4ca9ad55b75ff9a05debb36d5eb382e184 \\ \bottomrule
\end{tabular}
}
\label{tab:top10auction}
\end{table}

\subsection{Top-10 most spent addresses and top-10 name owners in Vickrey auction period.}

During Vickrey auction period, the addresses that spent most ETH and that owned most ENS names are shown in Table~\ref{tab:top10spent}.

\begin{table}[h]
\small
\centering
\caption{Top-10 most spent addresses and top-10 ENS name owners in the Vickrey auction period.}
\resizebox{\textwidth}{!}{
\begin{tabular}{@{}|crr|crr|@{}}
\toprule
Top-10 Most Spent Addresses in ETH &
  \multicolumn{1}{c}{\begin{tabular}[c]{@{}c@{}}Total \\ Sepnt \\ (ETH)\end{tabular}} &
  \multicolumn{1}{c|}{\begin{tabular}[c]{@{}c@{}}\# of \\ Names \\ Owned\end{tabular}} &
  Top-10 ENS Name Owner Addresses &
  \multicolumn{1}{c}{\begin{tabular}[c]{@{}c@{}}\# of \\ Names \\ Owned\end{tabular}} &
  \multicolumn{1}{c|}{\begin{tabular}[c]{@{}c@{}}Total \\ Spent \\ (ETH)\end{tabular}} \\ \midrule
0x8759b0b1d9cba80e3836228dfb982abaa2c48b97 & 39,712 & 32  & 0xa7f3659c53820346176f7e0e350780df304db179 & 26,257 & 1,686 \\
0x752975f5990c33da38c4cd50f0a41b70b3a6796c & 22,881 & 163 & 0x00bda1105ce38848b890c138d3d23a0435790a39 & 13,233 & 132   \\
0x9ef56cdfc51154a4bb25888a879560e08ad80795 & 13,073 & 215 & 0x001e28376ebe0982a50b0ad4a076a39aa0264bcc & 12,633 & 126   \\
0x352f25babf4a690673e35195efa8f79d05848aad & 7,928  & 56  & 0x00ab424d2019bc0f4c648232c6e7d181a34034b8 & 9,344  & 93    \\
0x217ad132e6271a9a19a818bde4cb6498888a3ba2 & 7,047  & 22  & 0x002acd20810b405fc4d01896871a6a7ba4b279fa & 5,481  & 55    \\
0xdc9fbececa49457fbcb2ee1dfd576a8be06a5c30 & 6,660  & 1   & 0xae18d320383a3598c65767dfd97c8df8ab465d26 & 4,627  & 47    \\
0x5807a8b404c71cf22eb0bac2e5f2a6c202ebe0a1 & 3,214  & 289 & 0xf5f700e1912b93ad09597bfa22484e01c0035b04 & 3,394  & 37    \\
0x00f2aaa26fa8a6aada2afa7f545b141c8aca983f & 3,139  & 65  & 0x000fb8369677b3065de5821a86bc9551d5e5eab9 & 3,071  & 31    \\
0xa37e710865314998ed47c3a24c8d1daaa58b1d07 & 2,695  & 8   & 0x64372db6405879214a0a76a7f1e9c013fd2fd84b & 2,307  & 23    \\
0xf14955b6f701a4bfd422dcc324cf1f4b5a466265 & 2,647  & 760 & 0x94048eb0db0ccb7d38219f28ed6522937b339aaf & 2,277  & 24    \\ \bottomrule
\end{tabular}
}
\label{tab:top10spent}
\end{table}

\section{Appendix for security issues and misbehaviors in Section~\ref{sec:security}}

\subsection{Top-10 holders of explicit squatting names.}

Table~\ref{tab:topalexaname} shows top-10 holders that registered most the known brands, which is a explicit squatting behavior.

\begin{table}[h]
\small
\centering
\caption{Top-10 holders of explicit squatting names.}
\begin{tabular}{@{}ccccc@{}}
\toprule
Address &
  \begin{tabular}[c]{@{}c@{}}\# of \\ Names \\ Owned\end{tabular} &
  \begin{tabular}[c]{@{}c@{}}Highest \\ Alexa \\ Rank\end{tabular} &
  Example &
  \begin{tabular}[c]{@{}c@{}}First \\ Registration \\ Time\end{tabular} \\ \midrule
0x009fde04525832da85a240d68c82421ca249a5b8 & 933 & 140  & ilovepdf.eth   & 2017/8/12 \\
0xf5f700e1912b93ad09597bfa22484e01c0035b04 & 478 & 23   & okezone.eth    & 2017/5/12 \\
0xa7f3659c53820346176f7e0e350780df304db179 & 262 & 425  & pearson.eth    & 2017/5/22 \\
0xaa0ea472b51ae4c5f77d50de5c91cbd932120b96 & 233 & 330  & gamepedia.eth  & 2017/9/3  \\
0xd2fa59b040852952bf4b4639edd4d8a718a4598a & 211 & 2053 & eonline.eth    & 2017/6/9  \\
0xc67247454e720328714c4e17bec7640572657bee & 206 & 662  & wiktionary.eth & 2017/5/20 \\

0xdc784458e2516a9d0531509f784dcf21983a60d7 & 184 & 162  & slideshare.eth & 2017/5/21 \\
0xd8c958f774de4b671e43f78fd0a04255e2291a13 & 160 & 34   & naver.eth      & 2017/6/12 \\
0x776efc479eeed9cb4fbc5e743aeb943313457bb5 & 153 & 358  & allegro.eth    & 2017/5/23 \\ 
0xf14955b6f701a4bfd422dcc324cf1f4b5a466265 & 136 & 35  & bongacams.eth & 2017/5/16 \\ \bottomrule
\end{tabular}
\label{tab:topalexaname}
\end{table}

\subsection{Top-10 holders of typo-squatting names.}
Figure~\ref{tab:toptyponame} shows the top-10 holders of typo-squatting names.

\begin{table}[]
\caption{Top-10 holders of typo-squatting names}
\begin{tabular}{@{}ccccc@{}}
\toprule
Address &
  \begin{tabular}[c]{@{}c@{}}\# of \\ Names \\ Owned\end{tabular} &
  \begin{tabular}[c]{@{}c@{}}Max \\ Alexa \\ Rank\end{tabular} &
  Example &
  \begin{tabular}[c]{@{}c@{}}First \\ Registration \\ Time\end{tabular} \\ \midrule
0xbd21109e2bdcb24c4fbcdc16a4c90f34e81228e2 & 1225 & 33  & zhuanqi.eth     & 2018/10/30 \\
0xa7f3659c53820346176f7e0e350780df304db179 & 948  & 70  & forbess.eth     & 2017/5/19  \\
0xbcbd4885ee8b2b74249c5ad9b8b668fb256a51b1 & 267  & 249 & brillo.eth      & 2017/5/13  \\
0x000fb8369677b3065de5821a86bc9551d5e5eab9 & 194  & 292 & allegri.eth     & 2017/5/15  \\
0xab48edd90bdf367d326d827758bacd2460c59d17 & 192  & 395 & hermès.eth      & 2019/9/29  \\
0x2bf1c1aedf56f2d1c413241d37f03c71b8793832 & 186  & 49  & live-jasmin.eth & 2017/6/10  \\
0x1d6f2f0356b3defadf14b1a0f8a3dcda89367d68 & 172  & 71  & facet.eth       & 2017/6/5   \\
0xddd3964d75d59b6b6d5c31eb313bba5ebf076364 & 159  & 378 & ubers.eth       & 2018/5/6   \\
0x813ec5facf289bd41365a8f1c9038a1228e95201 & 130  & 30  & bittrix.eth     & 2019/2/2   \\
0xf5f700e1912b93ad09597bfa22484e01c0035b04 & 123  & 1   & googlecom.eth   & 2017/5/26  \\ \bottomrule
\end{tabular}
\label{tab:toptyponame}
\end{table}

\subsection{Top-10 holders of ENS squatting names.}
After combining explicit ENS squatting names and typo-squatting names, the top-10 ENS squatting name holders are shown in Table~\ref{tab:topsushold}.

\begin{table}[h]
\small
\caption{The top-10 holders of ENS squatting names.}
\begin{tabular}{@{}cccc@{}}
\toprule
Address &
  \begin{tabular}[c]{@{}c@{}}Owned Squatting \\ Names (Unexpired)\end{tabular} &
  \begin{tabular}[c]{@{}c@{}}First \\ Registraion\end{tabular} &
  \begin{tabular}[c]{@{}c@{}}Owned Suspicious \\ Names (Unexpired)\end{tabular} \\ \midrule
0xbd21109e2bdcb24c4fbcdc16a4c90f34e81228e2 & 1276 (0)  & 2018/10/30 & 41370 (1)   \\
0xa7f3659c53820346176f7e0e350780df304db179 & 1210 (2)  & 2017/5/19  & 26199 (10)  \\
0x009fde04525832da85a240d68c82421ca249a5b8 & 974 (0)   & 2017/8/12  & 1491 (0)    \\
0xf5f700e1912b93ad09597bfa22484e01c0035b04 & 601 (0)   & 2017/5/12  & 3408 (0)    \\
0xbcbd4885ee8b2b74249c5ad9b8b668fb256a51b1 & 345 (340) & 2017/5/13  & 2366 (2262) \\
0xab48edd90bdf367d326d827758bacd2460c59d17 & 318 (318) & 2019/9/29  & 655 (655)   \\
0xdc784458e2516a9d0531509f784dcf21983a60d7 & 304 (174) & 2017/5/11  & 1306 (543)  \\
0xd2fa59b040852952bf4b4639edd4d8a718a4598a & 271 (1)   & 2017/6/9   & 2078 (1)    \\
0x2bf1c1aedf56f2d1c413241d37f03c71b8793832 & 259 (22)  & 2017/5/24  & 1532 (90)   \\
0xc67247454e720328714c4e17bec7640572657bee & 254 (4)   & 2017/5/20  & 1206 (13)   \\ \bottomrule
\end{tabular}
\label{tab:topsushold}
\end{table}

\subsection{Examples of websites with misbehaviors.}
\begin{figure}[h]
\centering
\includegraphics[width=0.9\textwidth]{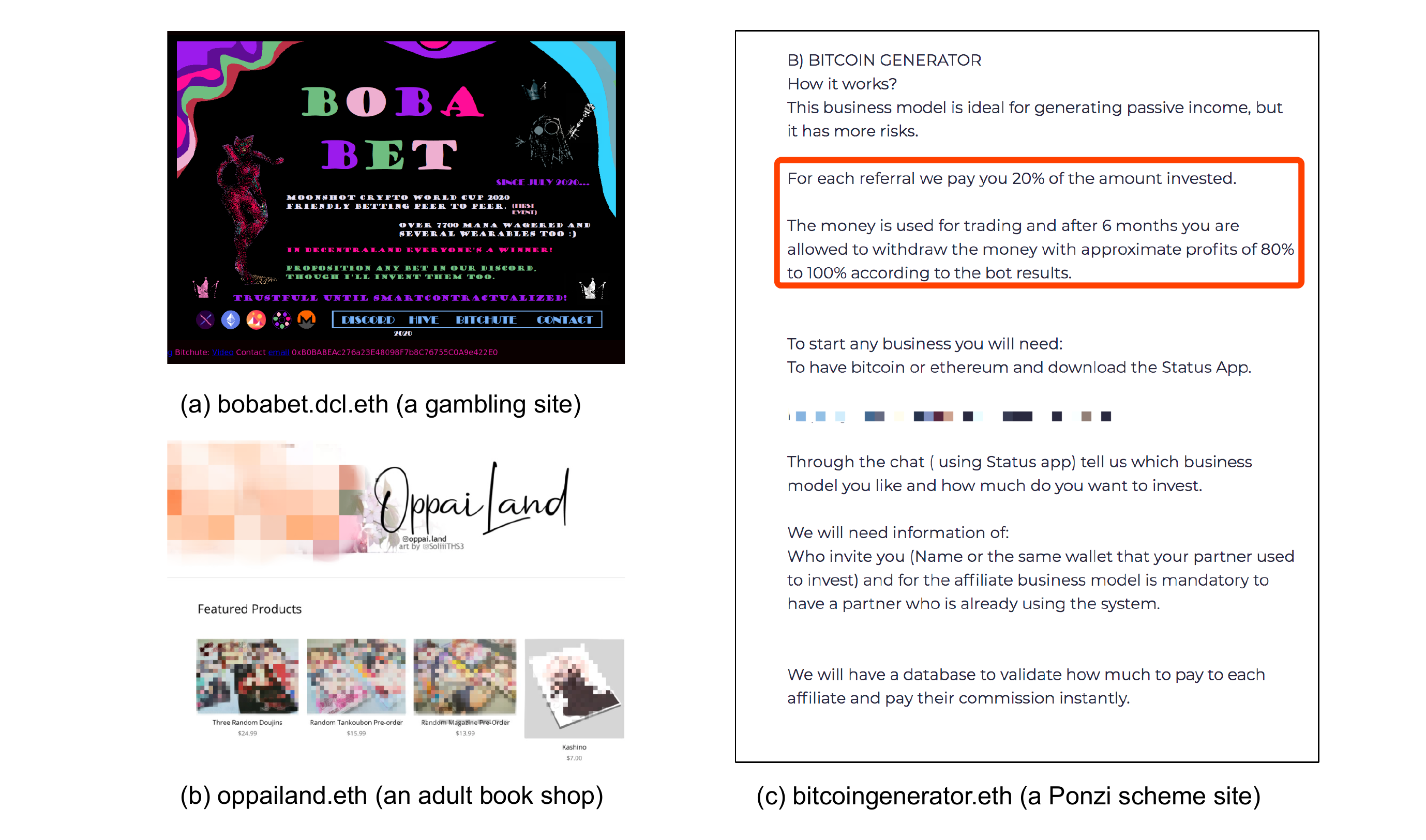}
\caption{Examples of websites with misbehaviors.}
\label{fig:scamdweb}
\end{figure}

Figure~\ref{fig:scamdweb} shows some examples of websites with misbehaviors we find in records of ENS.

\subsection{The screenshot of an airdrop scam using ENS.}
\begin{figure}[]
\centering
\includegraphics[width=0.45\textwidth]{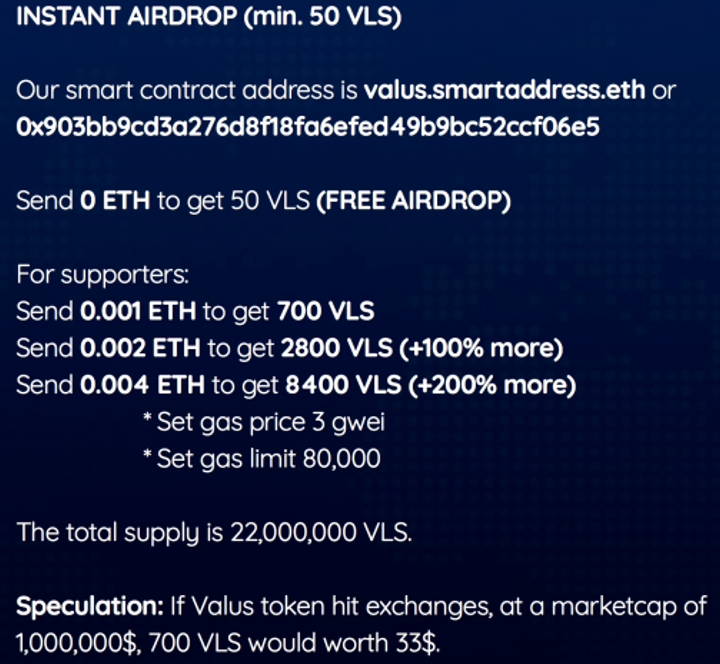}
\caption{The screenshot of airdrop scam using \texttt{valus.smartaddress.eth.}}
\label{fig:airdrop}
\end{figure}

Figure~\ref{fig:airdrop} shows the screenshot of an airdrop scam using \texttt{valus.smartaddress.eth}
 
\subsection{Some examples of expired (sub) domains with records.}
We have found 16,017 expired ENS names that have records in them or in their subdomains, examples of which are shown in Table~\ref{tab:expsubname}.

\begin{table}[h]
\small
\caption{Some examples of expired (sub) domains with records.}
\resizebox{0.7\linewidth}{!}{
\begin{tabular}{@{}|cc|ccc|@{}}
\toprule
.eth Name &
  \begin{tabular}[c]{@{}c@{}}Record \\ Type\end{tabular} &
  \begin{tabular}[c]{@{}c@{}}.eth Name with\\  Subdomains\end{tabular} &
  \begin{tabular}[c]{@{}c@{}}\# of \\ subdomains\end{tabular} &
  \begin{tabular}[c]{@{}c@{}}Record \\ Type\end{tabular} \\ \midrule
ammazon.eth    & ETH Address & thisisme.eth      & 708 & ETH Address \\ \midrule
wikipediaa.eth & ETH Address & tenzorum-id.eth   & 539 & ETH Address \\  \midrule
pay-pal.eth    & ETH Address & {[}unknown{]}.eth & 360 & Swarm Hash  \\ \midrule
investing.eth &
  \begin{tabular}[c]{@{}c@{}}ETH Address,\\ Swarm Hash\end{tabular} &
  portalid.eth &
  113 &
  \begin{tabular}[c]{@{}c@{}}ETH address,\\ Swarm hash,\\ Email\end{tabular} \\ \midrule
babytree.eth   & ETH Address & eth2phone.eth     & 61  & ETH address \\ \bottomrule
\end{tabular}
}
\label{tab:expsubname}
\end{table}


\end{document}